\documentclass[11pt]{article} 
 
\usepackage[margin=1in]{geometry} 
\usepackage{amsmath,amsthm,amssymb}
\usepackage{graphicx}
\usepackage{caption}
\usepackage{subcaption}
\usepackage[colorinlistoftodos]{todonotes}

\usepackage{natbib}
\setcitestyle{authoryear,open={(},close={)}}

\usepackage{booktabs}
\usepackage{multirow}
\usepackage{pdflscape}
\usepackage{titling}

\usepackage{etoolbox}
\apptocmd{\sloppy}{\hbadness 10000\relax}{}{}

\makeatletter
\DeclareRobustCommand{\textsupsub}[2]{{%
  \m@th\ensuremath{%
    ^{\mbox{\tiny#1}}%
    _{\mbox{\tiny#2}}%
  }%
}}
\makeatother

\usepackage{authblk}
\title{Reputation and Punishment sustain cooperation in the Optional Public Goods Game}
\author[1]{Shirsendu Podder}
\author[2]{Simone Righi}
\author[3]{Francesca Pancotto}
\affil[1]{Department of Computer Science, University College London, United Kingdom}
\affil[2]{Department of Economics, Ca'Foscari University of Venice, Italy}
\affil[3]{Department of Communication and Economics, Università degli Studi di Modena e Reggio Emilia}

\providecommand{\keywords}[1]
{
  \small	
  \textbf{\textit{Keywords---}} #1
}

\begin{document}
\date{}

\maketitle

\begin{abstract}
  Cooperative behaviour has been extensively studied as a choice between cooperation and defection. However, the possibility to not participate is also frequently available. This type of problem can be studied through the optional public goods game. The introduction of the ``Loner'' strategy, allows players to withdraw from the game, which leads to a cooperator-defector-loner cycle. While prosocial punishment can help increase cooperation, anti-social punishment -- where defectors punish cooperators -- causes its downfall in both experimental and theoretical studies.

  In this paper, we introduce social norms that allow agents to condition their behaviour to the reputation of their peers. We benchmark this both with respect to the standard optional public goods game and to the variant where all types of punishment are allowed. We find that a social norm imposing a more moderate reputational penalty for opting out than for defecting, increases cooperation. When, besides reputation, punishment is also possible, the two mechanisms work synergically under all social norms that do not assign to loners a strictly worse reputation than to defectors. Under this latter setup, the high levels of cooperation are sustained by conditional strategies, which largely reduce the use of pro-social punishment and almost completely eliminate anti-social punishment.
\end{abstract}


\clearpage
\tableofcontents
\pagebreak
\section{Introduction}

Explaining the conditions for the evolution of cooperation in groups of unrelated individuals is an important topic for natural and social scientists \citep{Axelrod1981,Rainey2003,Griffin2004,Nowak2006,Perc2017}. Research on the subject has been extensive and has lead to the identification of several mechanisms contributing to the success of cooperative behaviour: direct and indirect reciprocity, spatial selection, kin selection and multi-level selection, and punishment \citep{Nowak2006} among others. A large number of situations in which cooperation is difficult to achieve involve the pursuit of collective action to address problems beyond the individual dimension \citep{Smith1964}. Such problems are frequently studied through variations of the public goods game, where the public goods thrive through the contributions of its participants. While the social optimum is achieved when everyone's contribution is maximal, individuals always have the temptation to withhold their contributions, free-riding on the efforts of others \citep{Hardin1968}.

The public goods game has been mainly studied as a choice between contributing (or cooperating) and free-riding (or defecting) \citep{Ledyard1995}. However, in many situations, the possibility to not participate (or to exit) a situation is also available due, for example, to ostracism (\citealt{Maier-Rigaud2010}) or to individual choice (\citealt{Hauert2002}). An individual adopting this strategy chooses to abstain altogether from interactions, giving up not only its costs but also its potential benefits, instead preferring a lower but guaranteed payoff. For this reason, it is frequently called the ``Loner" strategy.

This behaviour has been observed in several social animal species.
When prides of lions hunt, individuals that actively pursue prey and contribute to their capture can be characterised as cooperators, members participating in the hunt but remaining immobile (or covering only very short distances) can be characterised as defectors, while the members hunting alone can be considered loners \citep{Scheel1991}. Similarly, groups of primates frequently display the practice of mutual grooming. In such situations, some individuals both actively receive grooming and groom others (a costly activity in terms of time and effort), while some fail to reciprocate the received benefits. Finally a third category of individuals tend to groom independently, refusing both to help and to accept help from others \citep{Smuts2008}. In humans, loners are those individuals that abstain from participation, from both the construction and from the benefits of the public good, for example refusing to participate in village activities involving conspicuous consumption \citep{Gell1986}, deciding not to join a business alliance, or quitting a secure job rather than intentionally reducing effort. Finally, at the institutional level, countries can decide to forgo joining geo-political alliances, rather than joining and then not engaging in their activities.

The forces influencing decision making in these situations can be captured in a game theoretical form through the Optional Public Goods Game (OPGG), which introduces the loner strategy into the traditional public goods game. This addition generates significant changes in the game dynamics, undermining the strength of defectors (who cannot exploit loners), while introducing cyclical population shifts. Indeed when most players cooperate, defection is the most effective strategy. When defectors become prevalent, it pays to stay out of the game and to resort to the loner strategy. Finally loner dominance lays the foundation for the return of cooperation \citep{Hauert2002,Hauert2002a}. Such {\it rock-paper-scissors} dynamics have been confirmed in experiments \citep{Semmann2003}. Relatedly, the possibility of opting out increases cooperation through enhanced prosociality of those who decide to stay in the game \citep{Orbell1993},  through the exit threat \citep{Nosenzo2017}, and through partner selection \citep{Hauk2003}.

Two mechanisms have been shown to be particularly important in influencing the ability of a group to engage in effective collective action: punishment and reputation. The former can be described as people willing to sustain an individual cost to punish behaviour that they find inappropriate \citep{Fehr2002}. Experiments \citep{Fehr2000, Henrich2006} have found that pro-social punishment (the punishment of a defector by a cooperator) -- whose emergence has been linked to the possibility to abstain from participation \citep{Hauert2007} -- can sustain cooperation in iterated games. More recently however, the positive role of punishment has been criticised, as substantial levels of anti-social punishment (the punishment of cooperators by defectors) have been observed in experiments \citep{Herrmann2008, Rand2010, Rand2011}. Although second-order free-riding on anti-social punishment can restore the effectiveness of pro-social punishment in the absence of loners \citep{Szolnoki2017}, allowing this type of punishment significantly reduces cooperation, and re-establishes the rock--paper--scissor type of dynamics among strategies \citep{Rand2011} in the OPGG. Importantly, \cite{Rand2011} show that the strategy of opting out of interactions, combined with a world in which anyone can punish anyone, does not result in any meaningful increase in cooperation. Crucially, it is not the presence of loners that harms cooperation, but the possibility to punish them. Indeed, if loners are shielded from punishment, \cite{Garcia2012} show that cooperators that punish pro-socially prevail, even when anti-social punishment is available. Just like anti-social punishment, anti-social pool rewarding (where agents of some type contribute towards rewarding others of their own type) also destabilises cooperation in fully-mixed populations \citep{Santos2015} but not structured populations \citep{Szolnoki2015}.

Besides punishment, the other key mechanism in sustaining cooperation towards collective action is reputation \citep{Milinski2002a}. Adopting a loose definition, reputation emerges when an individual's actions can be directly or indirectly observed by his peers and used to condition their own behaviour when playing with him. In both well-mixed \citep{Sigmund2001,Hauert2004a} and structured populations \citep{Brandt2003}, simple reputational systems promote and stabilise cooperation. Reputational systems at different levels of complexity have been studied, ranging from the first-order image scoring norm \citep{Nowak1998a}, to the second-order standing criterion \citep{Leimar2001,Panchanathan2003} and to the more complex third-order {\it leading eight} social norms \citep{Ohtsuki2004,Ohtsuki2006}. While simpler reputational systems have been observed in animals \citep{Bshary2006}, more elaborate systems are more likely to be the domain of human interactions, due to the complex relationship between actions, reputations and social structures.
How reputation evolves when a given behaviour is observed depends on the social norm characterising a population, in other terms their ``notion of goodness'', i.e. from the moral value attributed to a given type of action.

Punishment devices and reputation dynamics based on social norms coexist in human social groups \citep{Jordan2020} and therefore their interaction and implication for the success of collective action needs to be studied. In this paper, we extend the theory of OPGG studying the individual and joint impact of punishment and reputation on the sustainability of cooperation in a population where groups of agents engage in repeated interactions. In line with evidence that anti-social punishment exists and heavily influences cooperation \citep{Rand2011}, we allow for all types of punishment to occur, while studying its joint impact with a simple (first-order) social norm that prizes cooperation and penalises the loner and defection choices through bad reputations. By comparing the effect of social norms differing by the relative reputations they assign to defectors and loners, we explore the strategic exchanges between punishment and reputation. Similarly to \cite{Panchanathan2004}, we consider ephemeral groups of agents that interact repeatedly during each time period and are then reshuffled.

We find that while reputation alone mildly increases cooperation, and punishment alone does not, the two mechanisms synergistically interact leading to high cooperation. Our findings stem from the fact that reputation emerges as a partial substitute of punishment, thus making all types of punishment less necessary. Our results indicate a way through which social groups can sustain cooperation despite the presence of anti-social punishment.


\section{The Model}

A fully-mixed population of $N$ agents play the OPGG in randomly chosen groups of $n \geq 4$ as in \cite{Hauert2002}. During each interaction, players are given the choice between cooperating (C), defecting (D), or abstaining from the game (L). Cooperators incur a cost $c$ to participate and contribute towards the public good, while defectors participate but contribute nothing. Loners withdraw from the game to receive the loner's payoff $\sigma$ with $0 < \sigma < (r-1)c$. In each game, the sum of the contributions is multiplied by a factor $r > 1$ and is distributed equally between the group's participants. If there is only a single participant (or equivalently $n-1$ loners in the group), the OPGG is cancelled and all players are awarded $\sigma$. At this point, payoffs are awarded and additional games are played among the same agents until, with probability $1-\Omega$, the interactions terminate and the group is dissolved. Note that, unlike \cite{Milinski2002a}, our setup can explore a form of indirect reciprocity without the need of adding a different game with respect to the OPGG.

Following each round of the game, players have the opportunity to punish any or all of the other $n-1$ members of the group based on their played action. Players pay a cost of $\gamma$ for each player they choose to punish, while they are subjected to a penalty of $\beta$ for each punishment they receive.

Once the interaction and punishment rounds end, players change strategy based on the distribution of payoffs in the population. Each time-step an agent $i$ is randomly paired with another agent $j$ outside (within) its group with probability $m$ (respectively $1-m$). Player $i$ then imitates player $j$ with probability $\frac{\exp(u_j)}{\exp(u_i)+\exp(u_j)}$, where $u_x$ is the payoff of player $x$ and the exponential payoff function is used to account for negative utilities. Additionally, we allow mutations to occur in each period with probability $\epsilon$.

Extending the OPGG with the classical reputation mechanism of \cite{Brandt2003}, we implement a reputational system adopted by the entire population. After a player cooperates, defects, or abstains from the OPGG, he is assigned a reputation according to the population's social norm. A reputation system requires the definition of a social norm, or reputation dynamics \citep{Ohtsuki2004}, that identifies good and bad behaviour. The simplest possible reputational mechanisms are those that assign reputation on the basis of the mere observation of past actions, such as image scoring \citep{Nowak1998a}.
Arguably, most functional human and animal societies associate the highest reputation to cooperative behaviour. For this reason, we focus our attention to the reputation dynamics that prize cooperation above other actions. Like the binary image score, we simplify the reputation values so that players can have only a good (1), intermediate (0) or bad (-1) reputation, always assigning a good reputation to cooperators. We label the \textit{Anti-Defector (AD)} norm assigning reputations of -1 to defectors and 0 to loners. Similarly, the \textit{Anti-Loner (AL)} norm assigns loners a reputation of -1 and defectors a 0. Finally, it is possible to conceptualise norms that do not distinguish between acting as a loner or as a defector. We label these as the \textit{Anti-Neither (AN)} social norm, which assigns 0 to both actions, and the \textit{Anti-Both (AB)} social norm, which assigns -1 to both actions. Similar to punishment, reputations update after every OPGG round. We assume that actions are either observed without error or communicated honestly to everyone so that players' resulting reputations are common knowledge within the population.

To model agents adopting a social norm, we extend the set of behavioural strategies to allow for players to condition their actions on the average reputation of the other players within their group.
We condition actions on the average reputation firstly because a player's payoff is dependent on the actions of his group members and secondly because directly observing the actions of others can be difficult in sizeable groups, whereas observing the overall sentiment is arguably easier. Conditioning actions to the average reputation is also a key difference between our setup and \cite{Boyd1988} where actions depend on levels of cooperation, and from \cite{Takezawa2010}, where they depend on the amounts contributed. Indeed, given that there are three types of actions, each receiving a reputation on the basis of a shared social norm, reputational information can only be used as a signal of the cooperativeness of the environment a player is facing, similar to the case of image scoring. Furthermore, in our model reputation contains both elements of direct and indirect reciprocity as when groups are reshuffled, reputations are carried over. Finally, we significantly extend the strategy set considering strategies whose primary action in a benign  environment is not cooperating. The strategy X\textsupsub{$k_{\text{min}}$,Y}{$\text{Z}_\text{C}\text{Z}_\text{D}\text{Z}_\text{L}$} defines an agent taking action X $\in \{\text{C,D,L}\}$ if the average reputation of the other members of the group strictly exceeds $k_{\text{min}} \in [-1, 1)$, and Y $\in \{\text{C,D,L}\}$ otherwise. Furthermore, the subscripts describe the agent's strategy in terms of its punishment decision taken against cooperators Z\textsubscript{C}, defectors Z\textsubscript{D}, and loners Z\textsubscript{L}, where Z\textsubscript{C}, Z\textsubscript{D}, Z\textsubscript{L} $\in \{\text{P,N}\}$ is the decision to punish or not to punish respectively. In addition to the pure strategies of being an unconditional cooperator, defector and loner, this adds a further 8 strategies: C\textsupsub{-1,D}{}, C\textsupsub{0,D}{}, C\textsupsub{-1,L}{}, C\textsupsub{0,L}{}, D\textsupsub{-1,L}{}, D\textsupsub{0,L}{}, L\textsupsub{-1,D}{}, L\textsupsub{0,D}{}, each of which have 8 variants prescribing punishment towards zero, one or more cooperators, defectors and loners. In an effort to minimise the size of the strategy space, and therefore also some of the effects of random drift, we remove counterintuitive strategies, i.e. those that would cooperate only in groups of defectors or loners, defecting or going loner when surrounded by cooperators. For the same reason, strategies of higher complexity, i.e. those implying multiple behavioural shifts and/or more than two possible actions, are excluded from the present analysis and relegated to further studies.

Combining the payoffs of the game and the subsequent punishments, the utility of a player that decides not to opt out after a single round of the OPGG is:
\begin{equation*}
  u_i=  w_{i} - x_{i} + r\frac{\left(\sum_{j} x_{j} \right)}{n_{p}} - \beta \text{P} - \gamma \text{P}'
\end{equation*}
where the first two terms represent the initial endowment ($w_{i}$) and the initial contribution ($x_{i}\in\left\{0, c\right\}$) of the players. The third term represents the return from the OPGG (where $n_p$ is the number of agents that decided to participate in the game, i.e. to not opt-out), and the final two terms represent the losses incurred from punishing ($\beta$) and being punished ($\gamma$). In broad terms, $P$ represents the number of people in the group who exhibited the actions that player $i$ punishes, and $P'$ represents the number of people within the group that punish player $i$'s most recent action. For agents that decide to opt-out, the utility reduces to $u_i = w_i + \sigma - \beta \text{P} - \gamma \text{P}'$.
In summary, if we ignore reputation and punishment, we have 3 strategies, if we only have punishment, we have 24, if we only have reputation, we have 11, and if we have both reputation and punishment, we have 88 strategies.


\section{Results}

Investigating the dynamics of the OPGG (Fig. \ref{fig:action_distribution} and Fig. S\ref{fig:transition_matrix_baseline} in the Supplementary Materials), in the absence of both punishment and any reputational element, we reproduce the rock-paper-scissor dynamics between the three unconditional strategies as in \cite{Hauert2002} and \cite{Hauert2002a}. When aggregated over time, this results in agents cooperating or defecting about 20\% of the time each, while operating as loners about half of the time. Introducing the full set of punishing strategies as in \cite{Rand2011} we successfully reproduce their main result: the proportion of defectors and cooperators further diminishes with respect to the baseline, whilst the proportion of agents acting as loners increases. Having successfully replicated the key results of the extant literature, we investigate two models benchmarked respectively against the traditional OPGG and the OPGG with the full spectrum of punishment. In the first manipulation, we introduce a reputational system to the OPGG, while in the second we add it to the OPGG with punishment. In both cases we report results averaged over the second half of 100 simulations each comprising 200000 time steps.

\begin{figure}
  \centering
  \includegraphics[width=.75\textwidth]{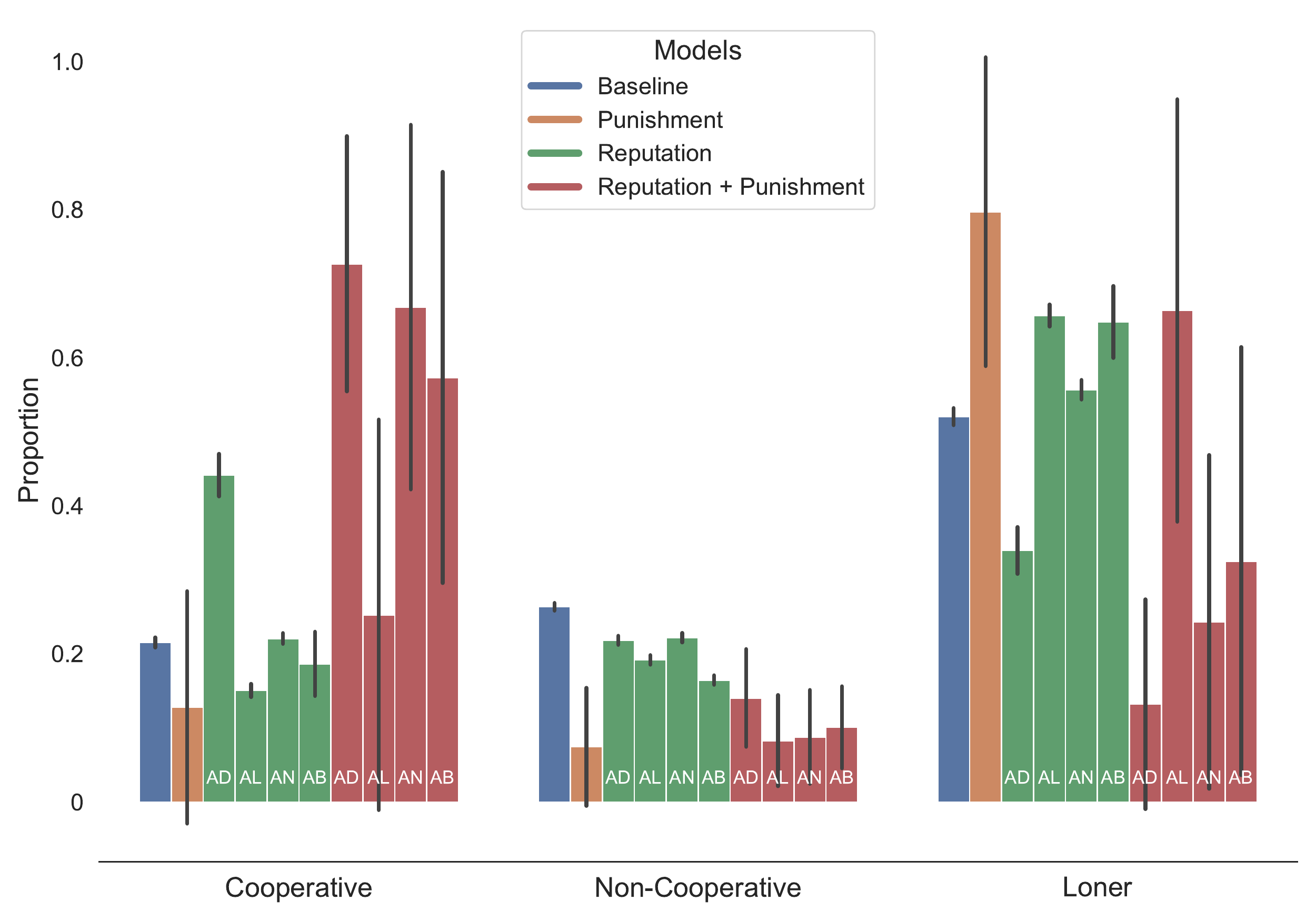}
  \caption{\textbf{Cooperation is highest when being a defector is strictly less reputable than being a loner, with and without punishment. Punishment and reputation together sustain high levels of cooperation as long as being a loner is not the strictly least reputable action. Without punishment, only the AD social norm sustains cooperation, with all other social norms providing comparable levels of (low) cooperation with the traditional OPGG in the absence of reputation and punishment. } All simulations shown use $N=1000,\, n=5,\, r=3,\, \sigma=1,\, m=0.95,\, \epsilon=0.1,\, \Omega=10/11$. Those including punishment additionally have $\gamma=1,\, \beta=2$. Results are calculated by averaging across the second half of the simulations, and then across 100 repeated simulations.}
  \label{fig:action_distribution}
\end{figure}

\begin{figure}
  \centering
  \includegraphics[width=\linewidth]{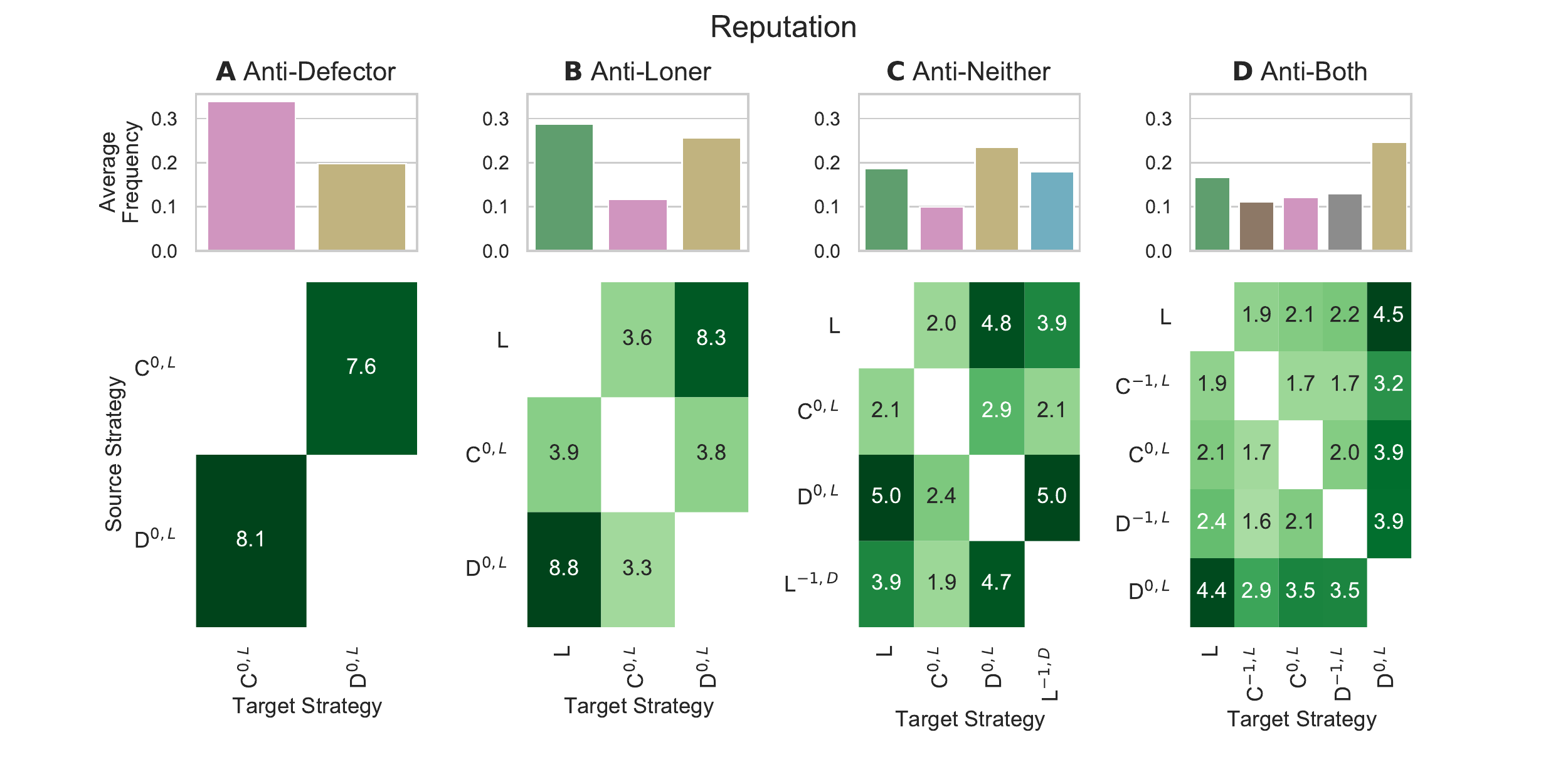}
  \caption{\textbf{In the absence of punishment, cooperation is sustained only when defection results in a strictly worse reputation than opting out. In this setup, a conditionally cooperating and a conditionally defecting strategy dominate the population.}}
  \label{fig:trans_matrices_condensed_rep}
\end{figure}

\begin{figure}[h]
  \centering
  \includegraphics[width=0.79\linewidth, trim=0 33 0 30, clip]{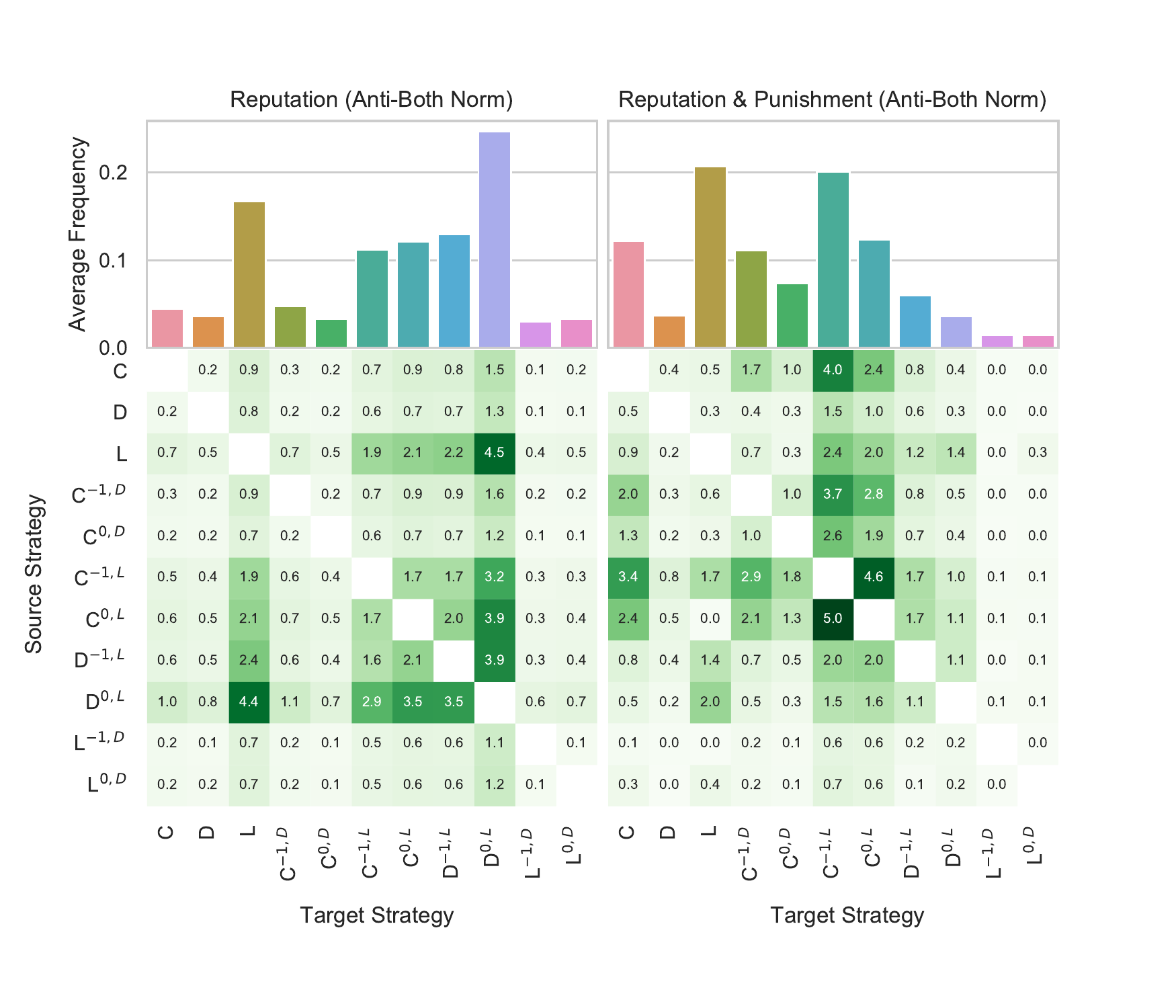}
  \caption{\textbf{Transition matrices with the  AB social norm, with (Right Panel) and without (Left Panel) punishment.} Like AD and AN, the AB norm displays high levels of cooperation when both reputation and punishment are used. Since under this norm, any behaviour other than cooperation results in a reputation assignment of -1, all conditional strategies that require a higher reputation threshold for their primary action are weakened.  C\textsupsub{0,D}{} and C\textsupsub{0,L}{} while C\textsupsub{-1,L}{} (and C\textsupsub{-1,D}{} to a slightly lesser extent) are the most popular. The world being largely populated by conditional cooperators and loners, causes the strategies C and L to be largely indistinguishable from their conditional variants, leading to a relatively large proportion of the population consisting of these unconditional strategies too. It is interesting to note that in the case with only reputation and the AB social norm, conditionally cooperative strategies (at least those who prefer to abstain than defect as their secondary action) are plentiful, yet the levels of cooperation are low (Fig. \ref{fig:action_distribution}) suggesting  that while a significant proportion of the population would like to cooperate, the environment is usually too harsh for them to risk it, so they act as loners. Parameters are identical to those in Fig. \ref{fig:action_distribution} averaged over 100 iterations. Numerical values in the matrices represent percentages of the total number of transitions in the entire population, excluding all transitions within the same strategy but to a different punishment variant.}
  \label{fig:transition_matrix_AB}
\end{figure}

\begin{figure}[!h]
  \centering
  \begin{subfigure}[b]{0.475\textwidth}
    \centering
    \includegraphics[width=\textwidth]{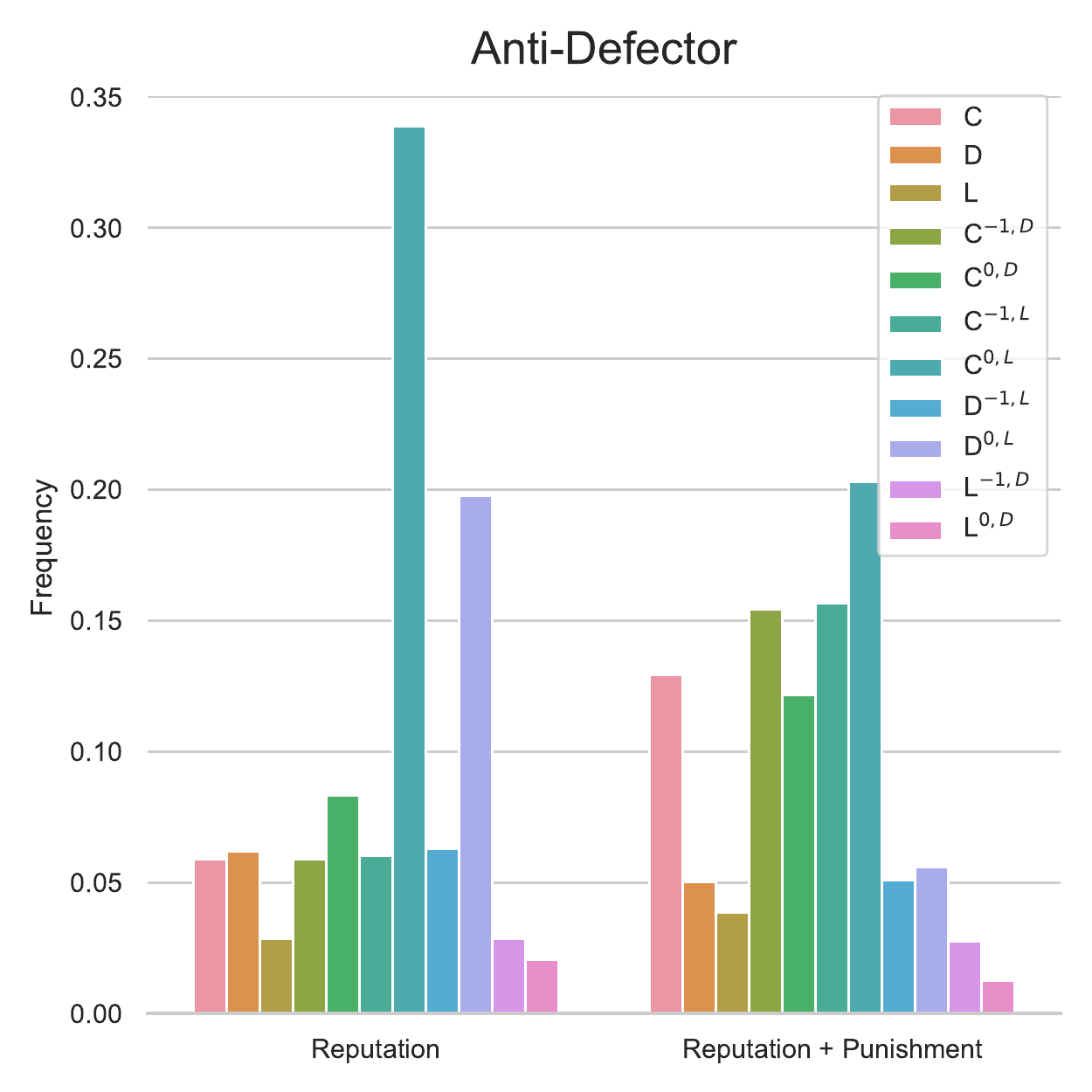}
  \end{subfigure}
  \hfill
  \begin{subfigure}[b]{0.475\textwidth}
    \centering
    \includegraphics[width=\textwidth]{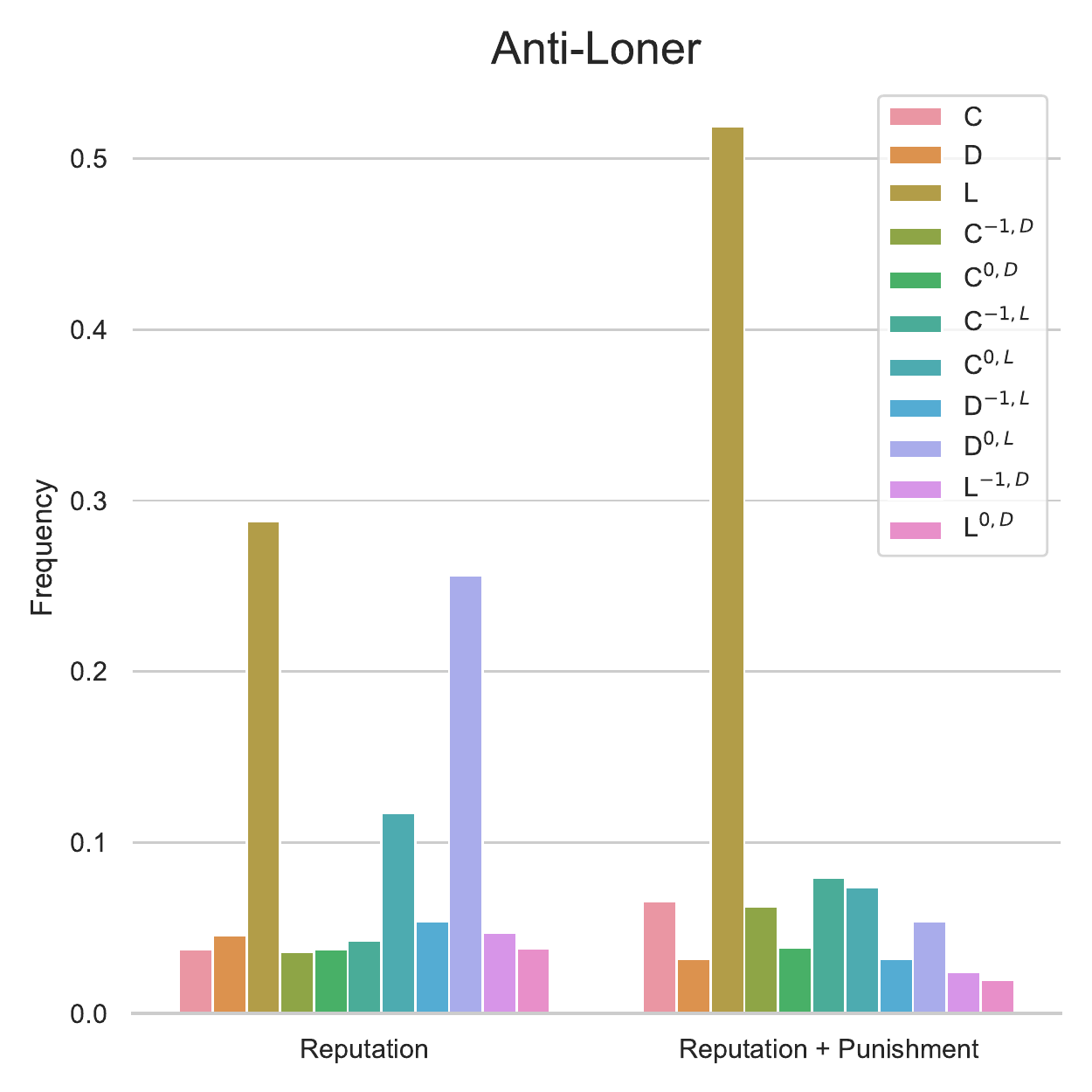}
  \end{subfigure}
  \vskip\baselineskip
  \begin{subfigure}[b]{0.475\textwidth}
    \centering
    \includegraphics[width=\textwidth]{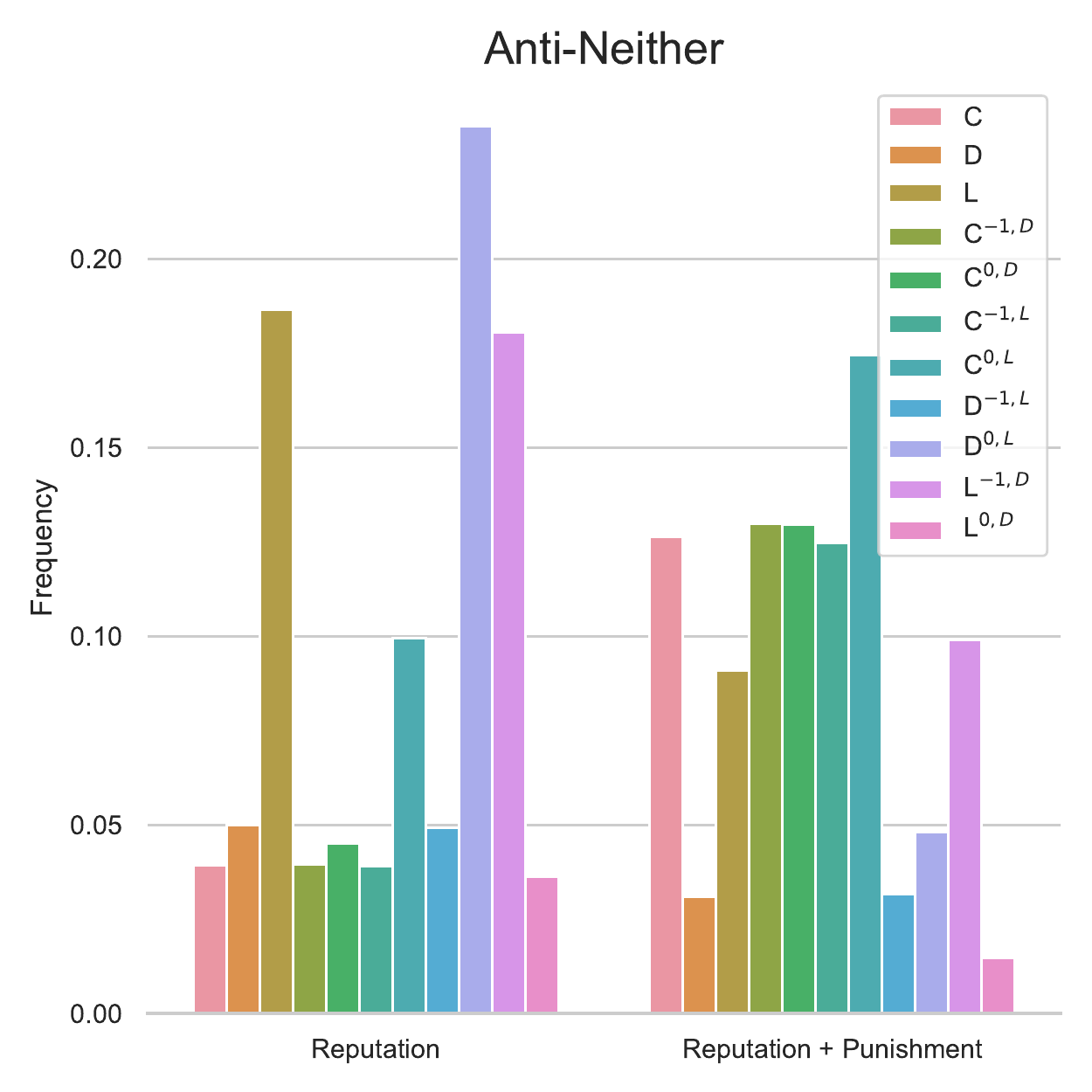}
  \end{subfigure}
  \hfill
  \begin{subfigure}[b]{0.475\textwidth}
    \centering
    \includegraphics[width=\textwidth]{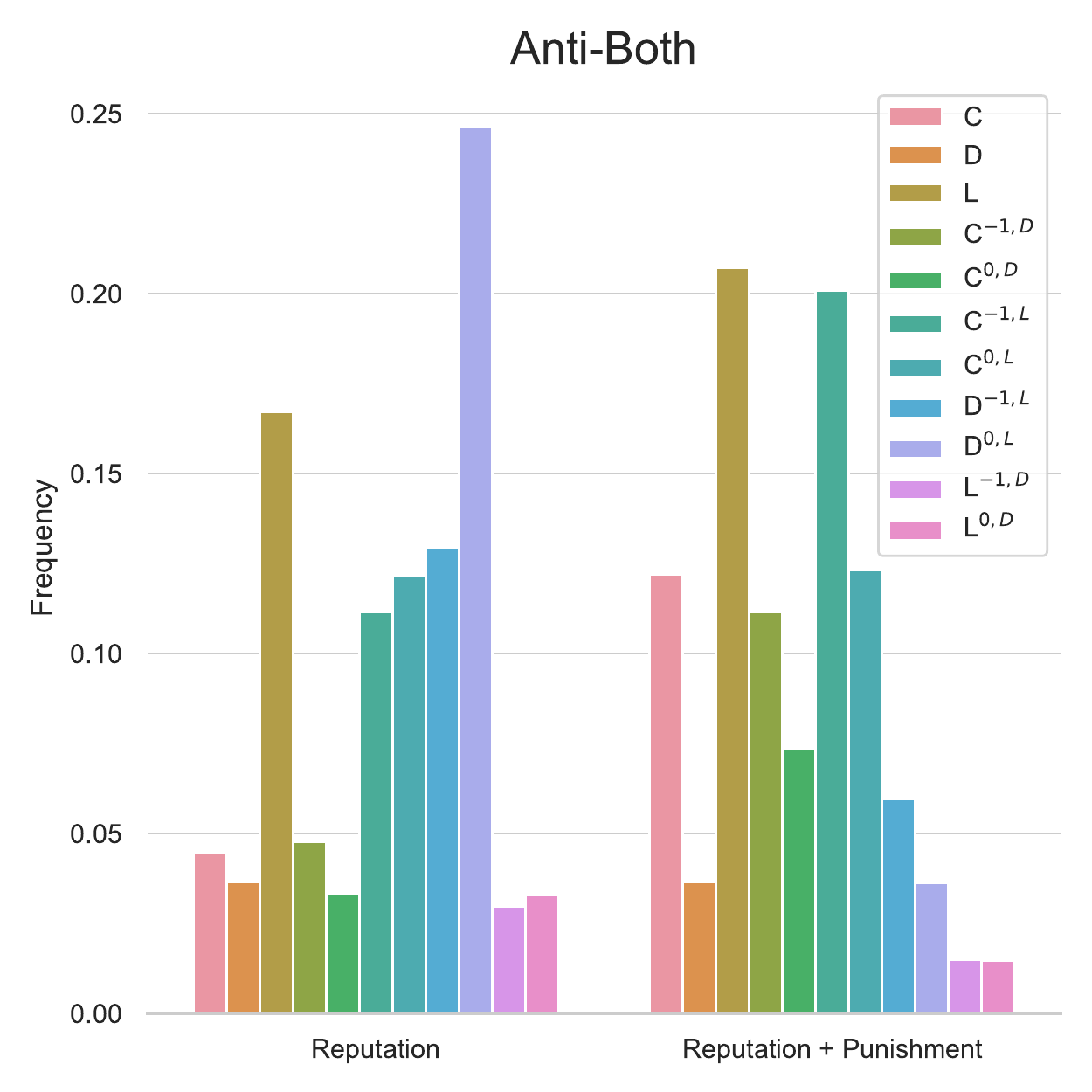}
  \end{subfigure}
  \caption{\textbf{Average population state under each model and social norm.} Each social norm has a different effect on the distribution of strategies within the population. The AD norm has the most favourable impact on cooperation, while the AL norm paradoxically emphasises the growth of loners. The AB norm improves the number of conditionally cooperative strategies within the population under solely reputation, but it is evident from Fig. \ref{fig:action_distribution} that they largely defect or opt-out as per their alternative directives. Both AB and AN norms provide the best chances for cooperation when populations use both reputation and punishment.}
  \label{fig:compositions_errorbars}
\end{figure}

\begin{figure}[!h]
  \centering
  \begin{subfigure}{0.48\textwidth}
    \includegraphics[width=1.1\textwidth]{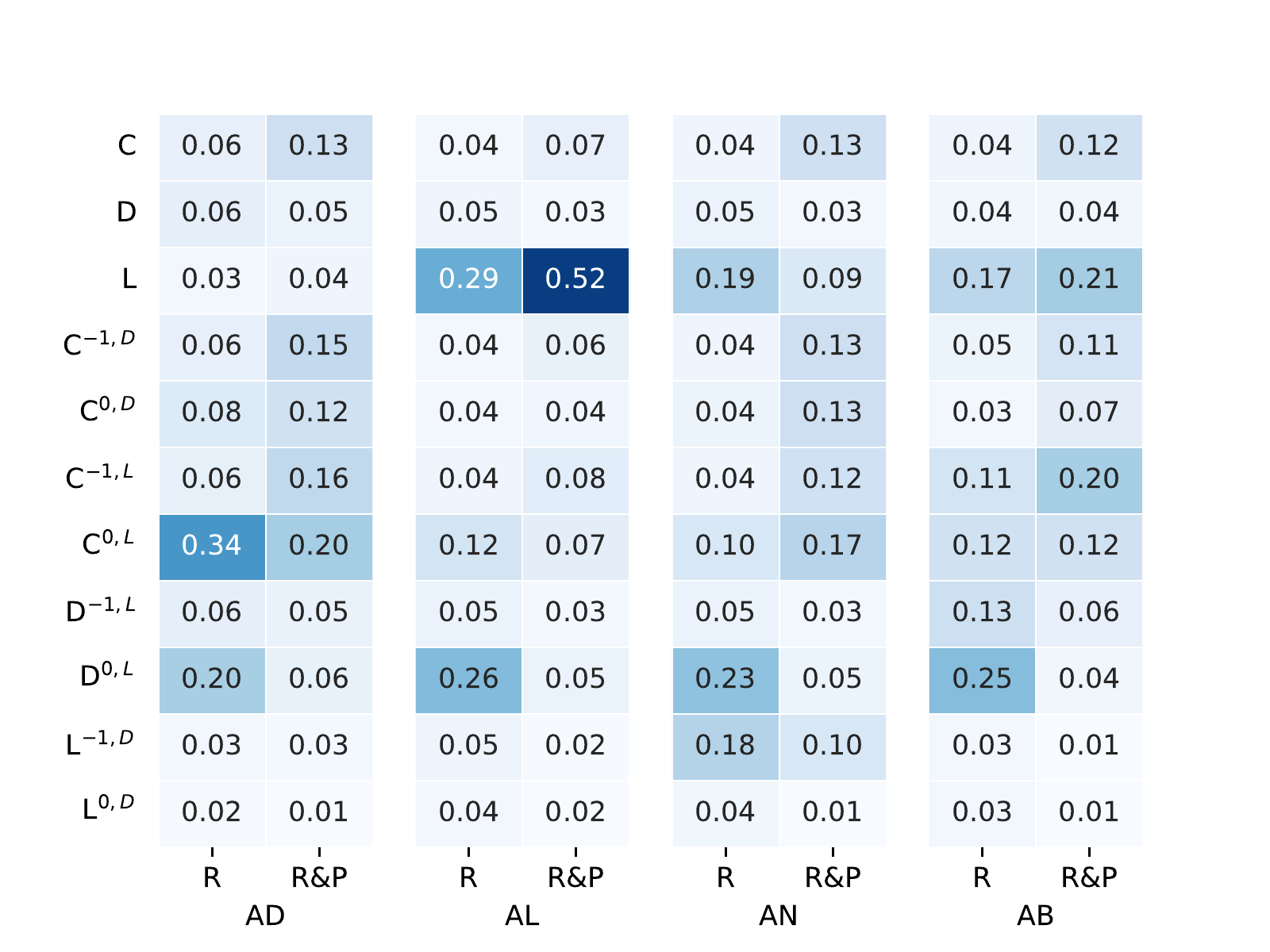}
    \caption{Average strategy proportions}
    \label{fig:compositions_a}
  \end{subfigure}
  \begin{subfigure}{0.48\textwidth}
    \includegraphics[width=1.1\textwidth]{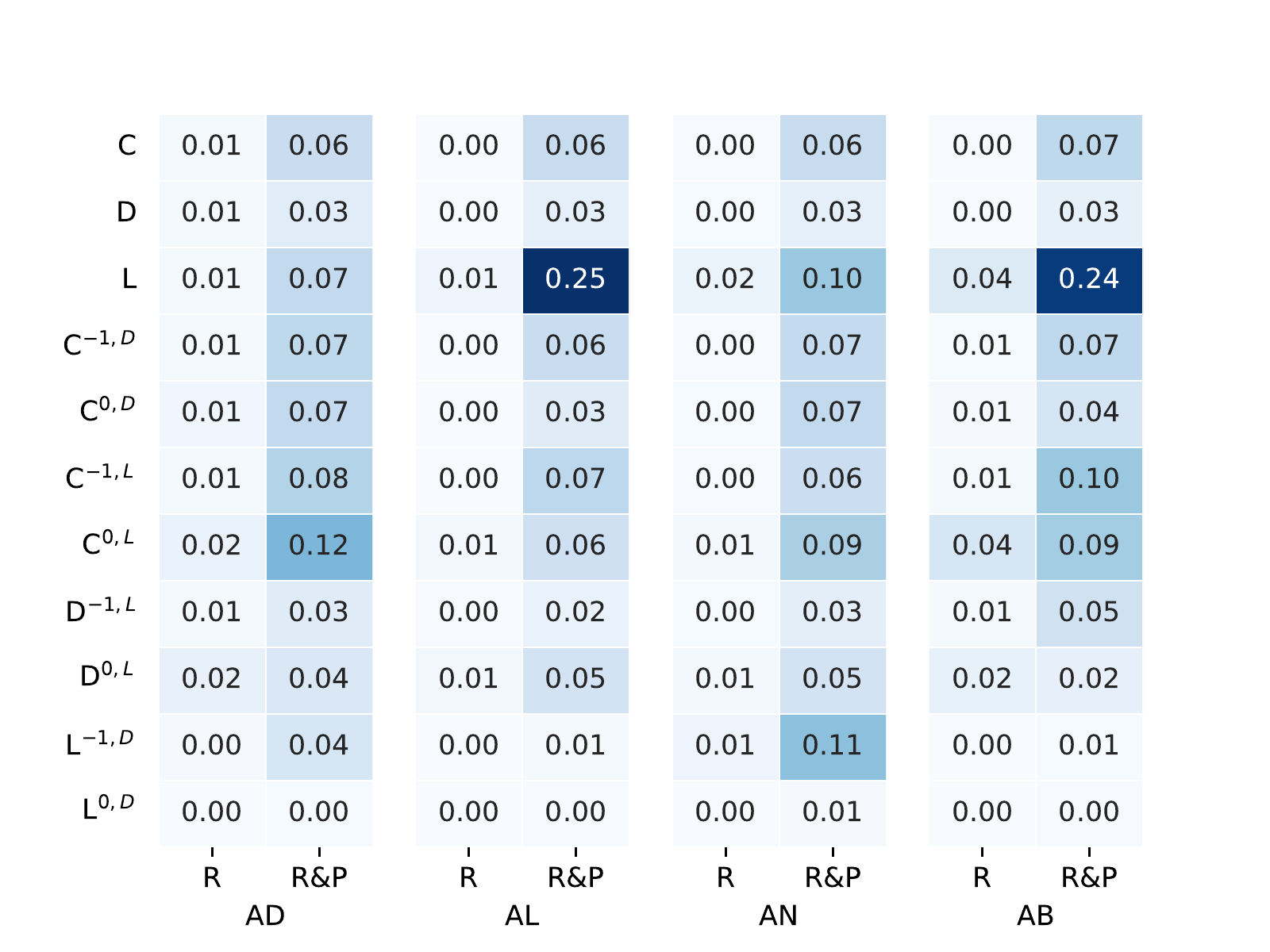}
    \caption{Standard Deviation of strategy proportions}
    \label{fig:compositions_b}
  \end{subfigure}
  \caption{\textbf{Mean (Left Panel) and standard deviations (Right Panel) of strategy compositions over each Model and Strategy Norm.} Columns titled R and R\&P represent the models utilising Reputation only and Reputation + Punishment respectively. In most cases, the distribution of strategies (when summed over their punishment variants) exhibit very low levels of variation. Populations that additionally implement punishment, exhibit larger standard deviations than those that do not.}
  \label{fig:compositions}
\end{figure}

\paragraph{Reputation without punishment}
We begin by analysing how the introduction of a simple reputational system influences the dynamics of these populations. The AD norm increases cooperation, so that in this circumstance, about 40\% of the OPGG actions were cooperative, roughly a 20\% increase from the baseline (Fig. \ref{fig:action_distribution}). Interestingly, the increase in cooperators does not come from the defector ranks, but rather from the loner ranks, which decrease the most with respect to the baseline. The increase in cooperation follows from a change in the patterns of strategy adoption caused by the introduction of reputation. Strategy adoption patterns are visualised, for each social norm studied, in Fig. \ref{fig:trans_matrices_condensed_rep} which also reports the reduced transition matrices (including only the strategies that are - on average - played by more than 10\% of the population). Results are computed averaging across time and then across 100 repetitions of each setup. The full population breakdown as well as the complete transition matrices are reported in the left panels of SM Figs. S\ref{fig:transition_matrix_AD} - S\ref{fig:transition_matrix_AB}. As noted in SM Figs. S\ref{fig:compositions_errorbars} and S\ref{fig:compositions}, the average population composition is remarkably stable across simulations, ensuring that the results are generalisable beyond a single simulation.

In the case of the AD social norm (that generates the highest levels of cooperation), pure strategies are largely abandoned in favour of conditional strategies. Among the 8 conditional strategies, only two are adopted by a sizeable part of the population: C\textsupsub{0,L}{} and D\textsupsub{0,L}{}.
These are strategies that share a common goal: to participate in interactions with a population of agents who are sufficiently cooperative (since they both set the reputational threshold for participation in their group to 0). However, while the former strategy wishes to engage with and enjoy the collaboration of their peers, the latter aims to exploit them.  When this is not possible, both strategies act as a loner for a lower but guaranteed payoff. From the transition table (SM Fig. S1, left panel), it emerges that most strategies turn into either C\textsupsub{0,L}{} or D\textsupsub{0,L}{}, and that these two strategies are much more likely to turn into each other than anything else (leftmost panel of Fig. \ref{fig:trans_matrices_condensed_rep}). This is a consequence of the interaction of these strategies and the level of cooperation in the population. This effect can be clearly discerned studying the temporal evolution of the proportion of agents playing these two strategies and of the associated actions played (Fig. S14, panel C). While the proportion of D\textsupsub{0,L}{} and C\textsupsub{0,L}{} in the population move systematically in opposite directions, studying the actual actions played we observe oscillations in the proportion of opt-out and cooperative actions, while defection is played by fewer individuals in a stable manner. In other terms the C\textsupsub{0,L}{} strategy uses the reputational information to conditionally cooperate. When in a group there happens to be too many defectors (most of whom are D\textsupsub{0,L}{}s given their relative prevalence) the strategy resorts to opting out, and thus avoiding being exploited. The strategy D\textsupsub{0,L}{} acts similarly, thus avoiding being completely evolved out, but it is unable to spread the defection in the population. Importantly, this dynamic relies on the fact that under the AD social norm, agents playing defection are assigned a strictly worse reputation than loners, thus allowing C\textsupsub{0,L}{} to discriminate between the two behaviours. In other terms, the cyclic behaviour of the baseline model - described by \cite{Rand2011} - reemerges, but with the loner strategy being part of two more complex conditional strategies. Importantly, the cycles are - in this setup - quite noisy due to the occasional attractiveness (Fig. S\ref{fig:transition_matrix_AD}, left panel) of other strategies, which allows them to be the destination of some transitions and thus to survive albeit in low numbers.

\begin{figure}[!h]
  \centering
  \begin{subfigure}[b]{0.64\textwidth}
    \centering
    \includegraphics[width=\textwidth]{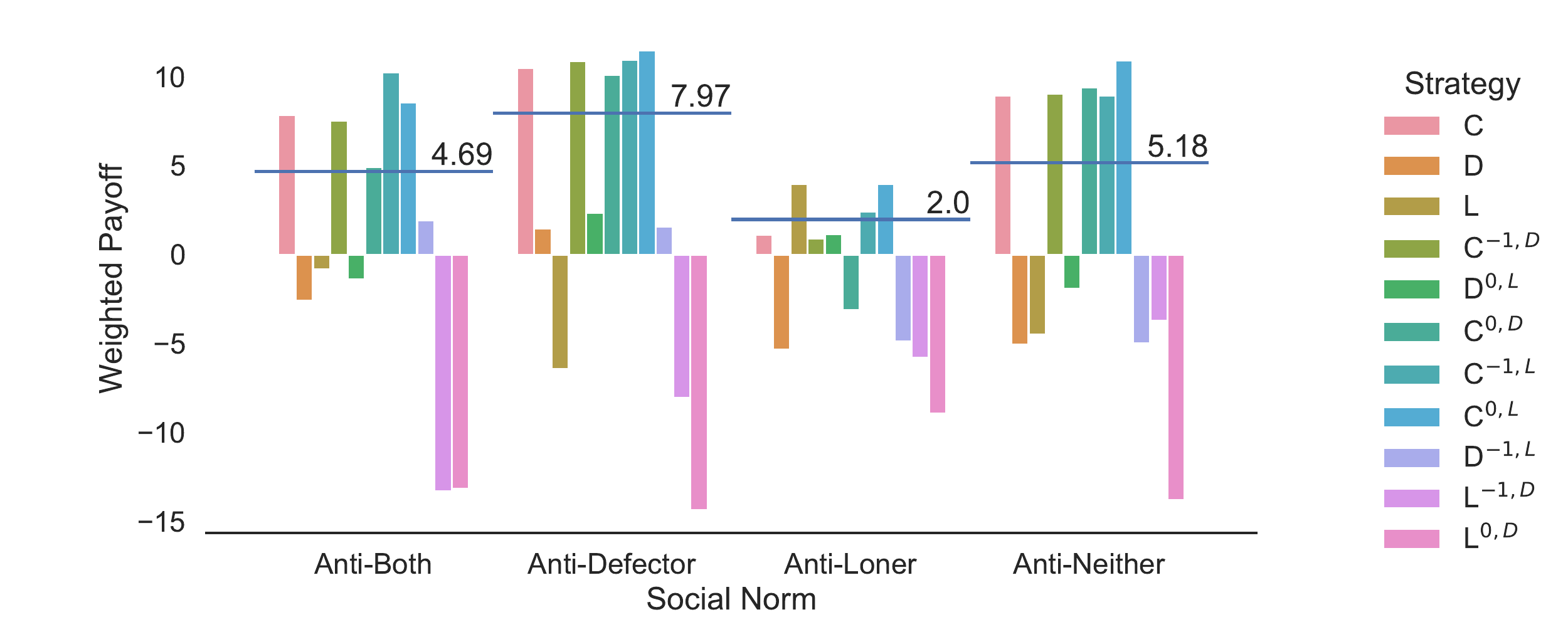}
    \caption{Reputation and Punishment}
  \end{subfigure}
  \vskip\baselineskip
  \begin{subfigure}[b]{0.64\textwidth}
    \centering
    \includegraphics[width=\textwidth]{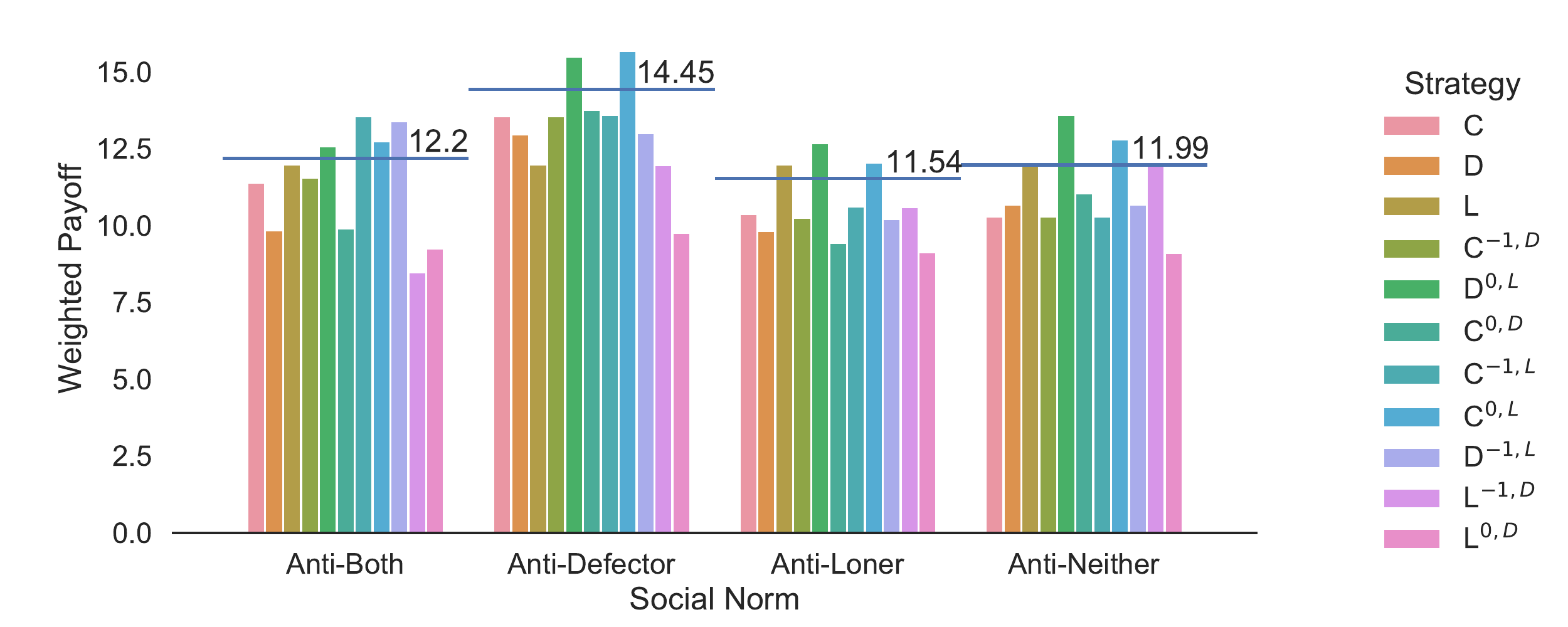}
    \caption{Reputation}
  \end{subfigure}
  \vskip\baselineskip
  \begin{subfigure}[b]{0.36\textwidth}
    \centering
    \includegraphics[width=\textwidth]{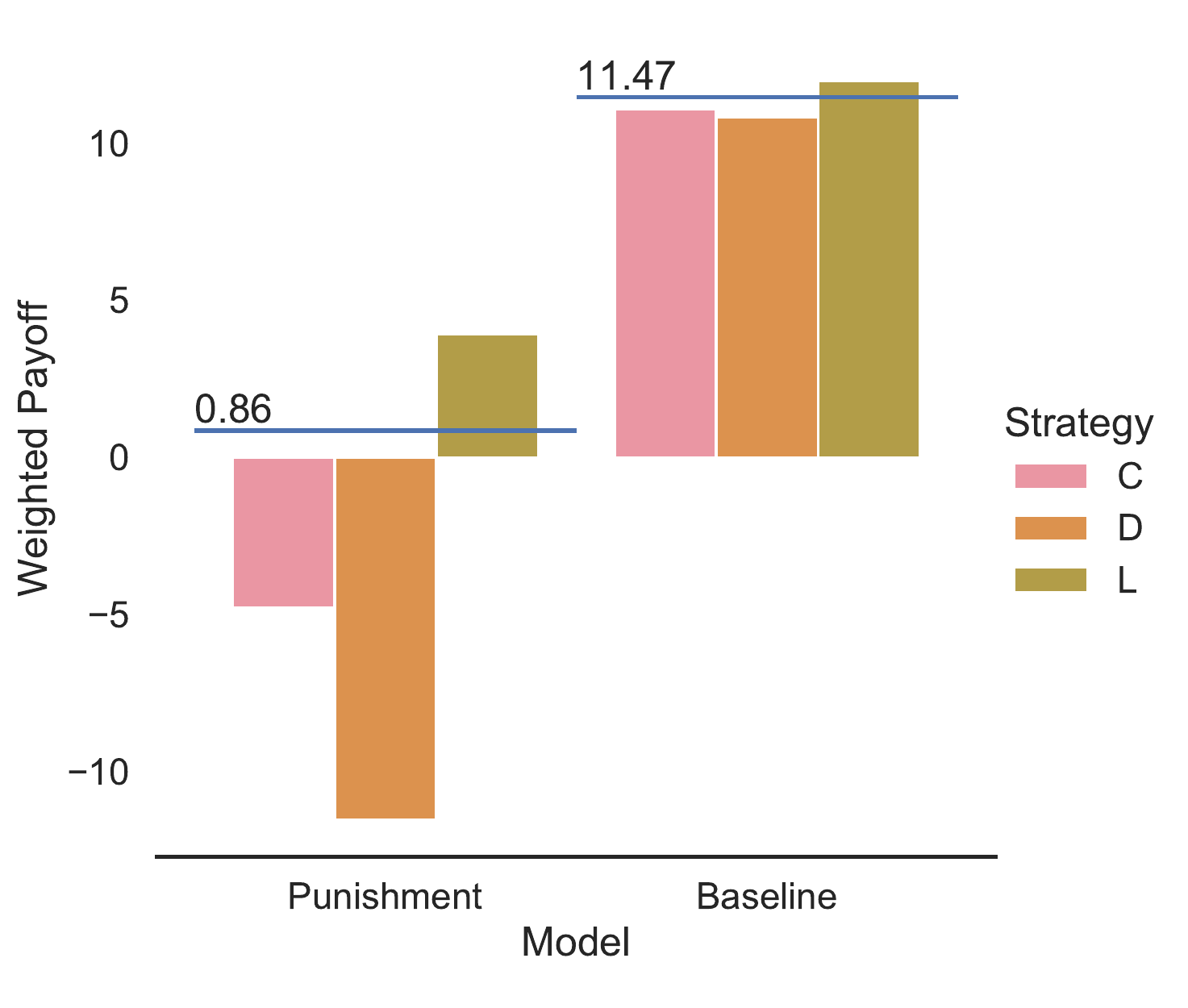}
    \caption{OPGG with and without punishment}
  \end{subfigure}
  \caption{\textbf{Average payoffs of each strategy in each model. The AD norm is the most profitable with or without punishment.} In all social norms, regardless of whether or not punishment is present, the best chances for cooperation emerge under the Anti-Defector social norm. The Anti-Loner social norm is the most harmful towards population fitness. When both reputation and punishment are at play (Panel (a)), the AN norm is preferable to the AB norm, most likely due to assigning a lower reputation towards loners, allowing more protection from exploitative behaviour. However, in the absence of punishment (Panel (b)), the AB norm is slightly more conducive towards cooperation than AN. We reproduce the baseline results of the traditional OPGG and the OPGG with all punishment strategies in Panel (c). }
  \label{fig:payoffs}
\end{figure}

\begin{figure}[h]
  \centering
  \includegraphics[width=0.65\linewidth, trim=0 5 0 23, clip]{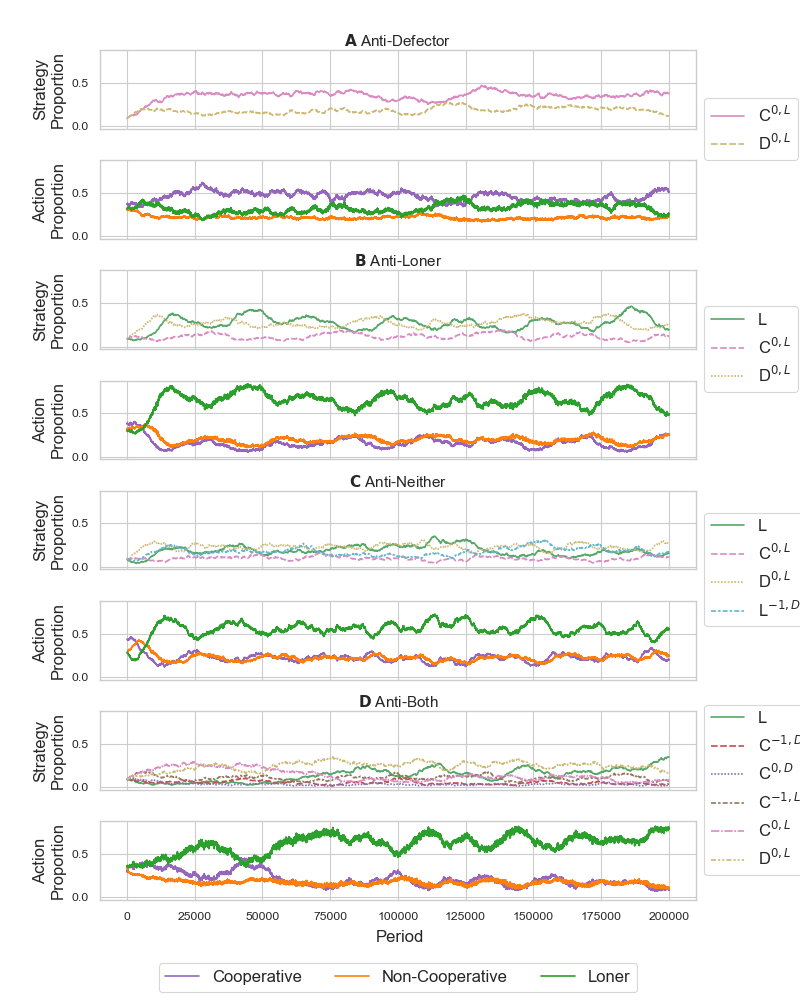}
  \caption{\textbf{Time series representations of singular simulations, reputation only.} Panel A shows the time evolution of strategies under the AD social norm. This is the only social norm that sustains any cooperation in the absence of punishment, with a stable level of low defection. Panels B-D show populations where the only safe action to take is to abstain and while periods in which loners are the majority are followed by a small increase in cooperation, this is soon lost as defectors take advantage. The social norms that attribute the worst reputation to being a loner have the greatest levels of players abstaining.}
  \label{fig:series_rep}
\end{figure}

The peculiarity of AD is immediately evident when studying the other social norms. While the anti-defector social norm results in both the greatest population fitness (Fig. S\ref{fig:payoffs}) and increase in cooperation, all other social norms are unable to sustain levels of cooperation higher than the baseline, resulting in populations largely dominated by loners (Fig. \ref{fig:action_distribution}) with a cyclical pattern similar to those of  \cite{Rand2011}.  The reason for this divergence lays in the reputational damage inflicted by these norms to non-cooperators. Indeed while reputation helps agents condition their behaviour to opponents' types, attributing loners a bad reputation does not inflict on them any direct damage. Since they are already prone to not participating in the game, their payoffs remain independent from any reputational effect. However, attributing loners too low a reputation (in AL and AB) hampers the role of loners in providing a backup option against invading defectors. Furthermore, not attributing defectors a strictly lower reputation (in AN), allows them to more easily exploit cooperators by impeding the latter to discriminate between loners and defectors (the full transition matrices are reported in SM Figs. S\ref{fig:transition_matrix_AD} - S\ref{fig:transition_matrix_AB}). In all these cases the effectiveness of C\textsupsub{0,L}{} against defection is more limited hence the strategy has limited evolutionary success (Fig. \ref{fig:trans_matrices_condensed_rep}, panels B, C, D and SM Fig. S\ref{fig:series_rep}). For this reason, in the following we will focus our attention mainly on the AD social norm.

\begin{figure}
  \centering
  \includegraphics[width=\linewidth]{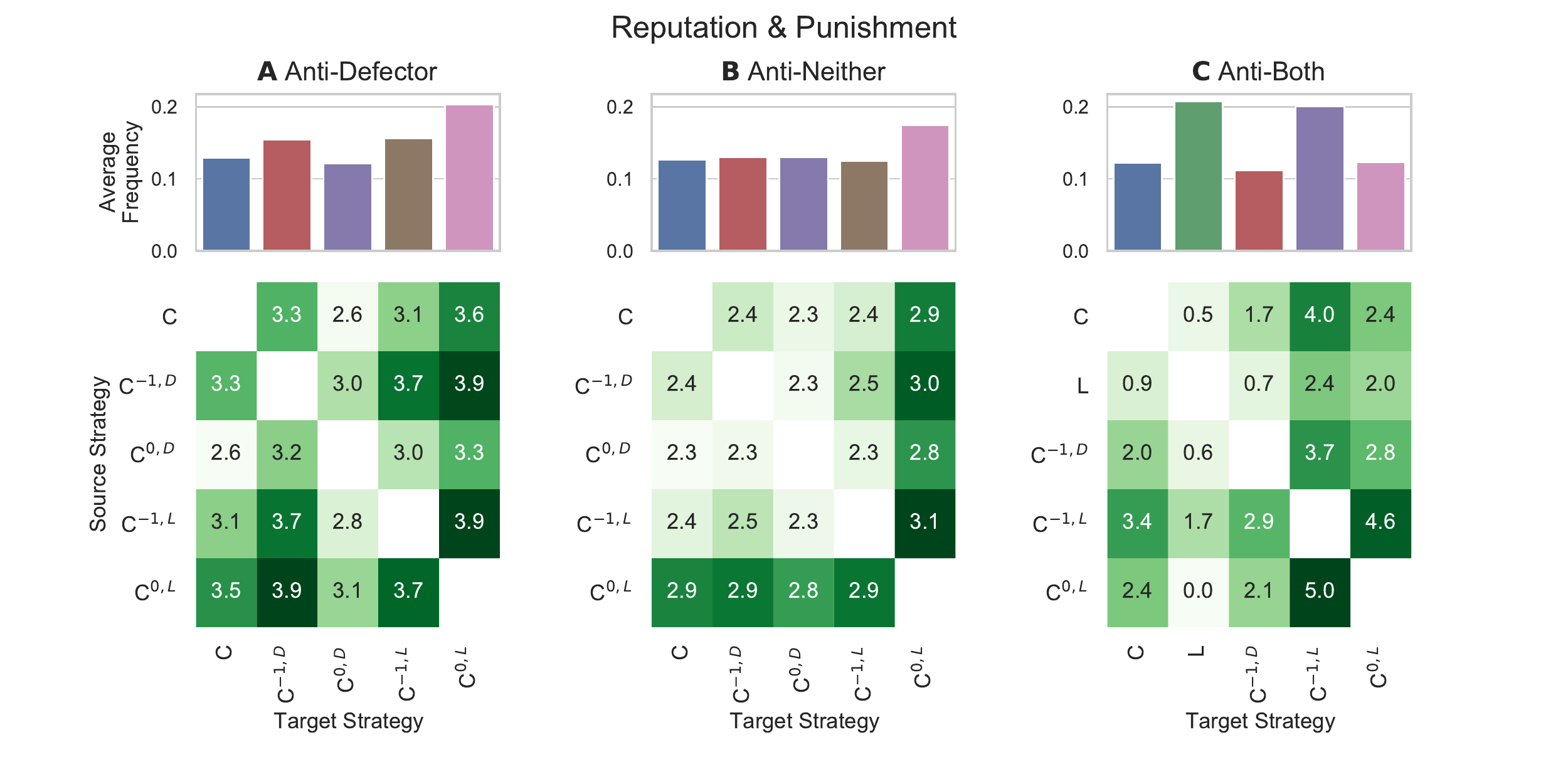}
  \caption{\textbf{When punishment is available, high levels of cooperation evolve as long as being a loner is not the strictly least reputable action. Strategies that thrive in these situations conditionally cooperate, becoming indistinguishable between one other in such a cooperative environment.} Under the AB norm, both defecting and exiting are the least reputable actions which creates both the most hostile environments and the least reputable OPGG groups. This results in conditional cooperators being cautious by frequently opting out. The AL social norm is not pictured as the only long-term dominant strategy that exists in significant amounts is the loner strategy. See SM Figs. S\ref{fig:transition_matrix_AD}-S\ref{fig:transition_matrix_AB} for the complete transition matrices.}
  \label{fig:trans_matrices_condensed_reppun}
\end{figure}

\begin{figure}
  \centering
  \includegraphics[]{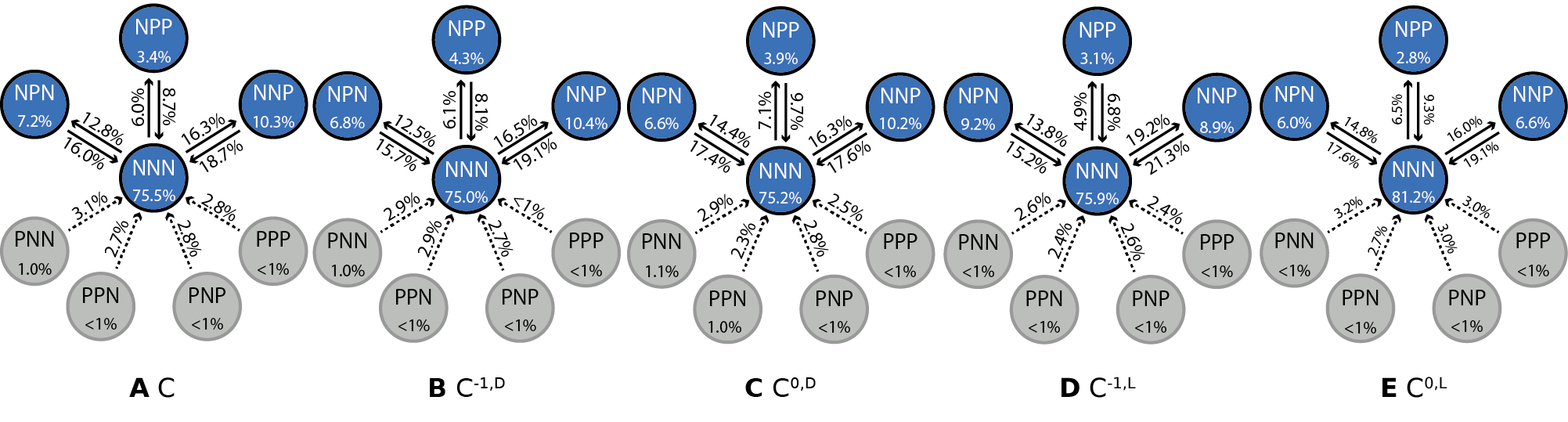}
  \caption{\textbf{When both punishment and reputation are available, most individuals do not punish, a few punish pro-socially, and almost no one punishes anti-socially.} Nodes represent the punishment variants of each of the 5 most popular strategies within the reputation and punishment model (AD social norm). The values within the nodes are the average proportion of the variant relative to all variants of the strategy. Edges represent the transitions solely between variants, normalised by the total number of transitions between all punishment variants of the strategy. The strategies that do not punish are always the most popular of the variations, those that punish some combination of defectors or loners are common, while those that punish cooperators never survive for long. See SM Fig. S\ref{fig:intrastrategy_rep_pun_transitionmatrix_C} for more details.}
  \label{fig:transition_network_internal}
\end{figure}

\paragraph{Reputation with punishment}
We now analyse the impact of introducing the AD norm to the OPGG with punishment. Due to the substantial empirical and theoretical evidence for the presence of anti-social punishment in the OPGG, we allow all possible types of punishment to occur. In the presence of the AD norm, the level of cooperation in the population increases significantly with cooperation becoming the most played strategy, at the expense again of loners whose role becomes marginal in this setup. The level of cooperation is not only higher than when solely punishment is used, but is also of the case where only reputation is available. Given the success of this cooperation, one could expect punishment and the reputation system to interact in some form, coevolving to produce successful strategies that both punish pro-socially and cooperate conditionally. However, this is largely not the case. Analysing the prevalence of strategies and the transition probabilities between them (Fig. \ref{fig:trans_matrices_condensed_reppun} for between strategy transitions, and Fig. \ref{fig:transition_network_internal} for transitions between different punishment patterns for the same basic strategy), we note that -- for each of the conditional strategies that become dominant in the population -- the variants which employ punishment are quite rare. When punishers are born through mutation, they rapidly turn into their non-punishing counterparts.

\begin{figure}[h!]
  \centering
  \includegraphics[scale=0.75]{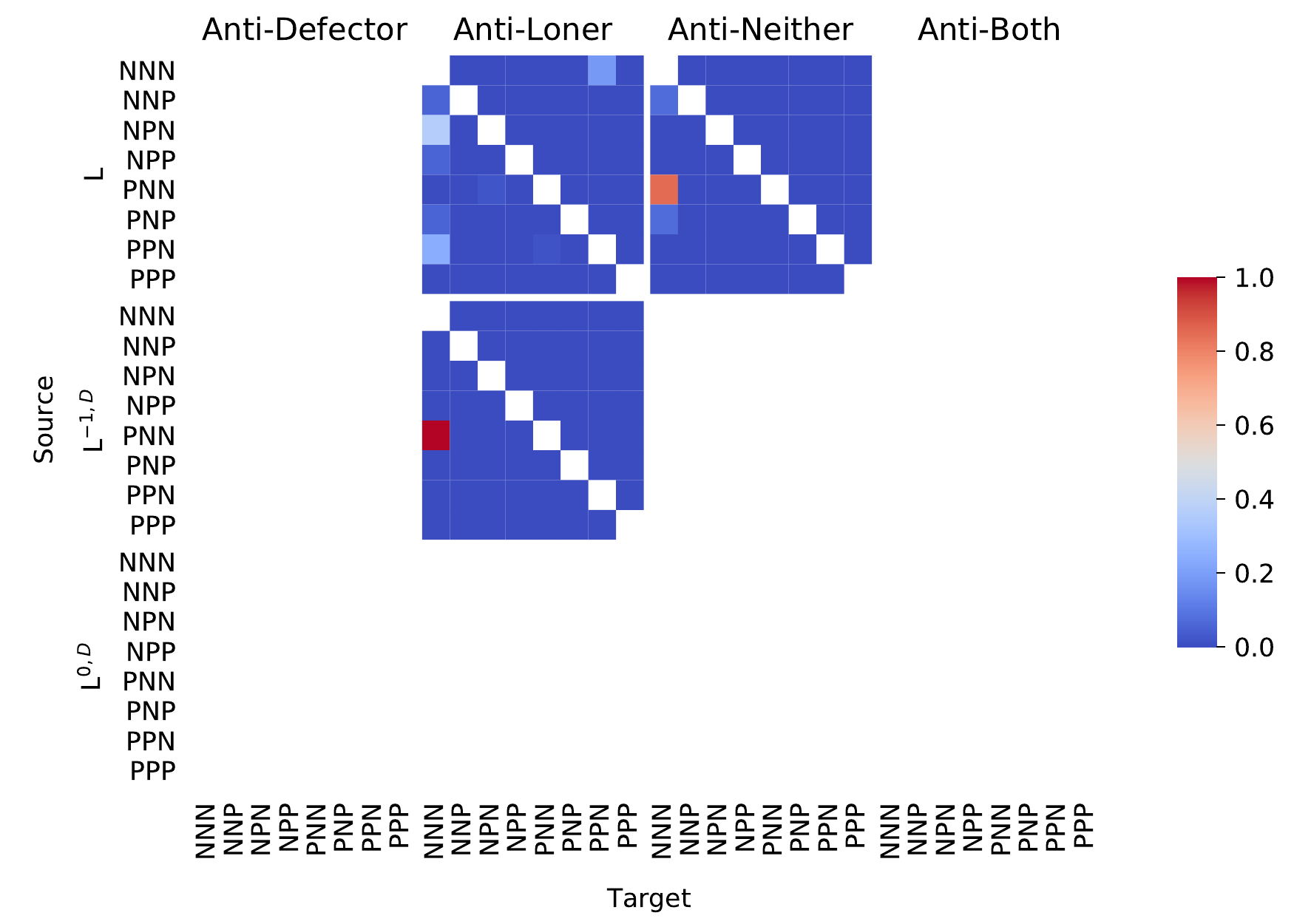}
  \caption{\textbf{Transition matrices between punishment variants of loner strategies.} The conditional strategies that primarily opt-out of the OPGG tend to have a very weak presence within the population (Fig. \ref{fig:compositions}). As such, any transitions within variants of the same strategy tend to occur only very rarely. This suggests that upon introduction to the population, either at the beginning of the simulation or through mutation, loner strategies immediately transition away to a strategy with completely different behaviour, hence the lack of transitions for several strategies and social norms in this figure. Simulation parameters are identical to those in Fig. \ref{fig:action_distribution} with 100 iterations per parameterisation. The colour intensity represents the movement between punishment variants, relative to the total number of transitions between variants of that strategy.}
  \label{fig:intrastrategy_rep_pun_transitionmatrix_L}
\end{figure}

While representing consistently low proportions of the population, the types of punishment that show more resilience (being sometimes able to attract non-punishing counterparts), are those that punish defectors and/or loners, i.e. the pro-social types (see SM Figs. S\ref{fig:transition_matrix_AD} - S\ref{fig:intrastrategy_rep_pun_transitionmatrix_L} for the complete transition matrices and the prevalence of each strategy within each model and social norm). Thus when both punishment and reputation exist, the latter acts as a substitute for the former. Agents evolve toward strategies that do not punish and away from strategies that do, contributing to the demise of this kind of behaviour, except as a threat against deviators.

Regardless of punishment patterns, the conditional strategies that emerge as prevalent are those whose primary action is to cooperate, which in turn provides the ideal environment for unconditional cooperators to thrive. Indeed,  the combination of punishment and reputation ensures that agents that adopt strategies whose primary action is different from cooperation, end up with lower average payoffs given that the established cooperative conditional strategies will be less likely to cooperate with individuals whose primary action is not cooperation. The transition matrix of Fig. \ref{fig:trans_matrices_condensed_reppun}, shows that the surviving strategies  shift between each other without a clear pattern. This is likely due to fluctuations in the group composition when playing the OPGG as these strategies are almost equivalent in largely cooperative environments.

\begin{figure}[!h]
  \centering
  \begin{subfigure}[b]{\textwidth}
    \centering
    \includegraphics[width=0.8\textwidth]{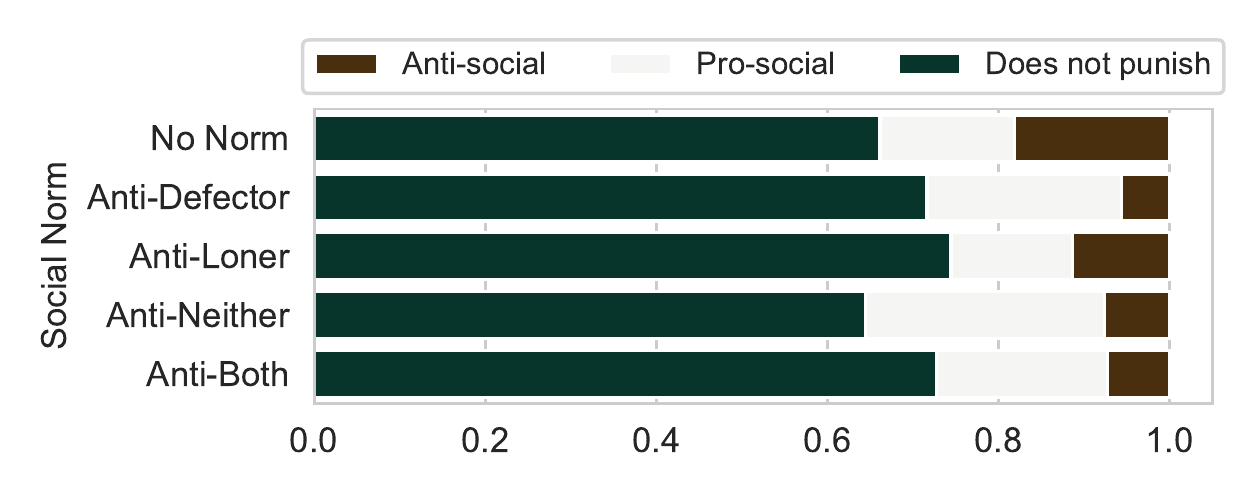}
    \caption{Reputation and Punishment - grouped by type of punishment}
  \end{subfigure}%
  \vfill
  \begin{subfigure}[b]{0.8\textwidth}
    \centering
    \includegraphics[width=\textwidth]{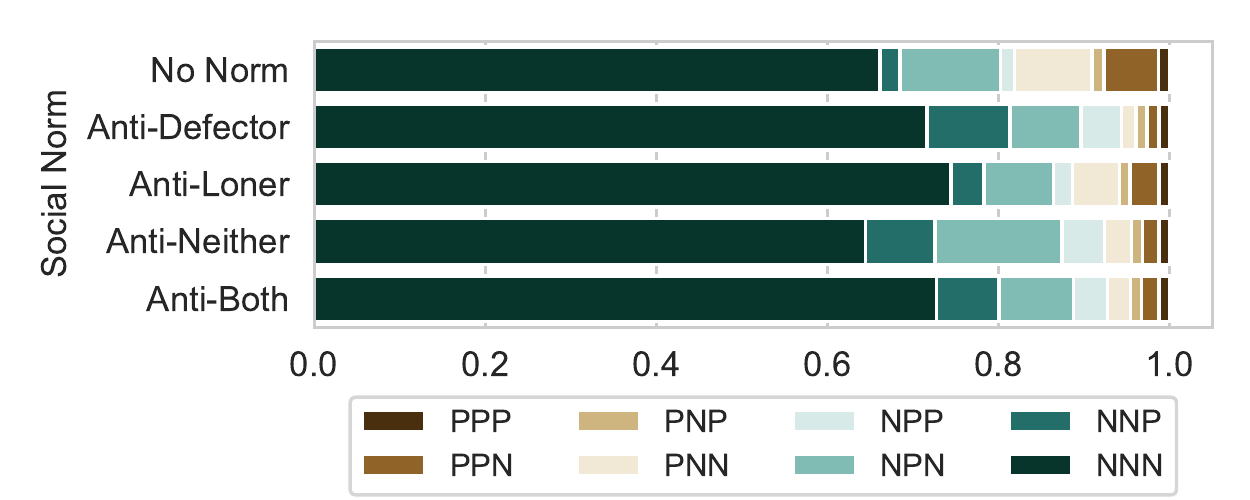}
    \caption{Reputation and Punishment - grouped by punishment variant}
  \end{subfigure}
  \caption{\textbf{Social norms categorised by punishment-type proportions.} \emph{Panel (a):} All social norms lower the levels of anti-social punishment in the population, and with the exception of the AL norm, increase pro-social punishment relative to the overall level of punishment in the population. \emph{Panel (b):} The pro-social punishment of defectors (NPN, NNP and NPP) essentially doubles in AN compared to the AD norm, confirming our finding that costly punishment compensates for weaker sanctioning through social norms. Comparing the AD and AB social norms, assigning loners a worse reputation makes little difference, confirming our view that since Loners are largely independent, they remain indifferent to others' perceptions towards them. Parameters are identical to those of Fig. \ref{fig:action_distribution} averaged over 100 iterations.}
  \label{fig:punishment_overview}
\end{figure}

\begin{figure}[!h]
  \centering
  \begin{subfigure}[b]{0.43\textwidth}
    \centering
    \includegraphics[width=0.9\textwidth]{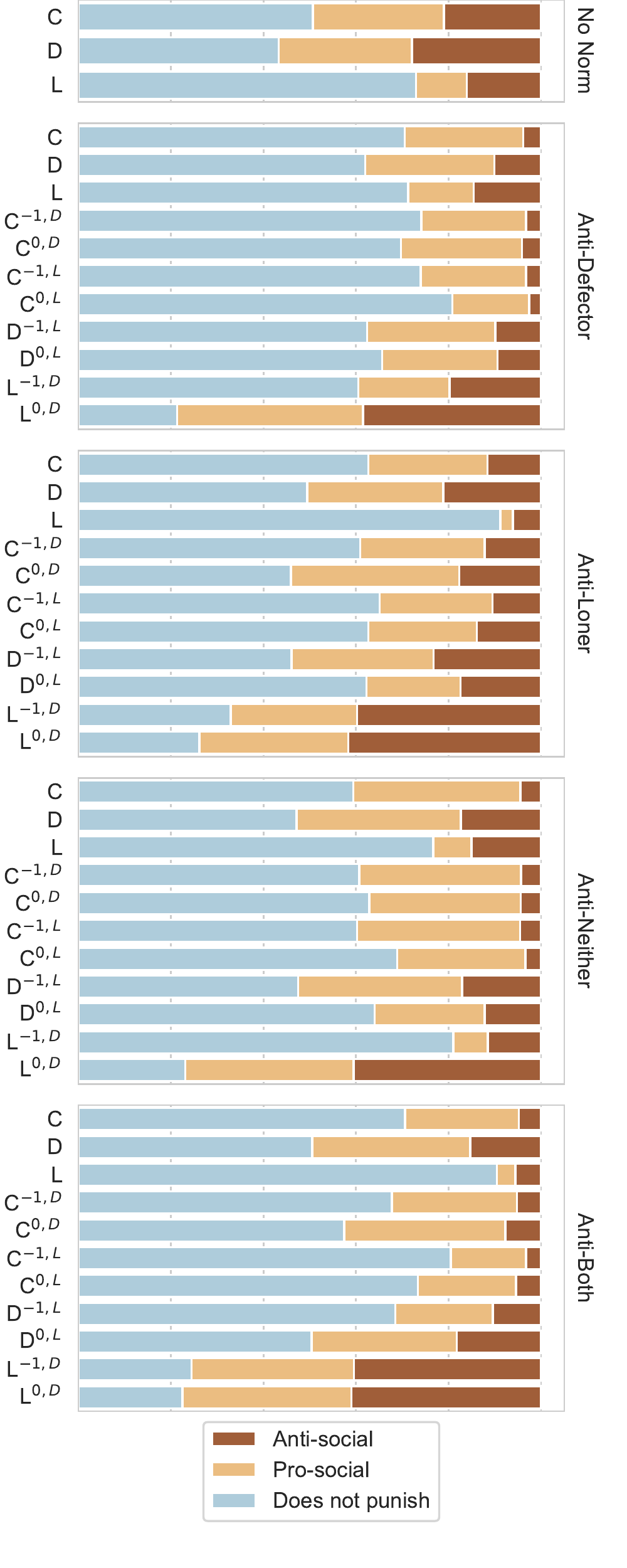}
    \caption{Reputation and Punishment - grouped by strategy and type of punishment}
  \end{subfigure}%
  \hfill
  \begin{subfigure}[b]{0.43\textwidth}
    \centering
    \includegraphics[width=0.9\textwidth]{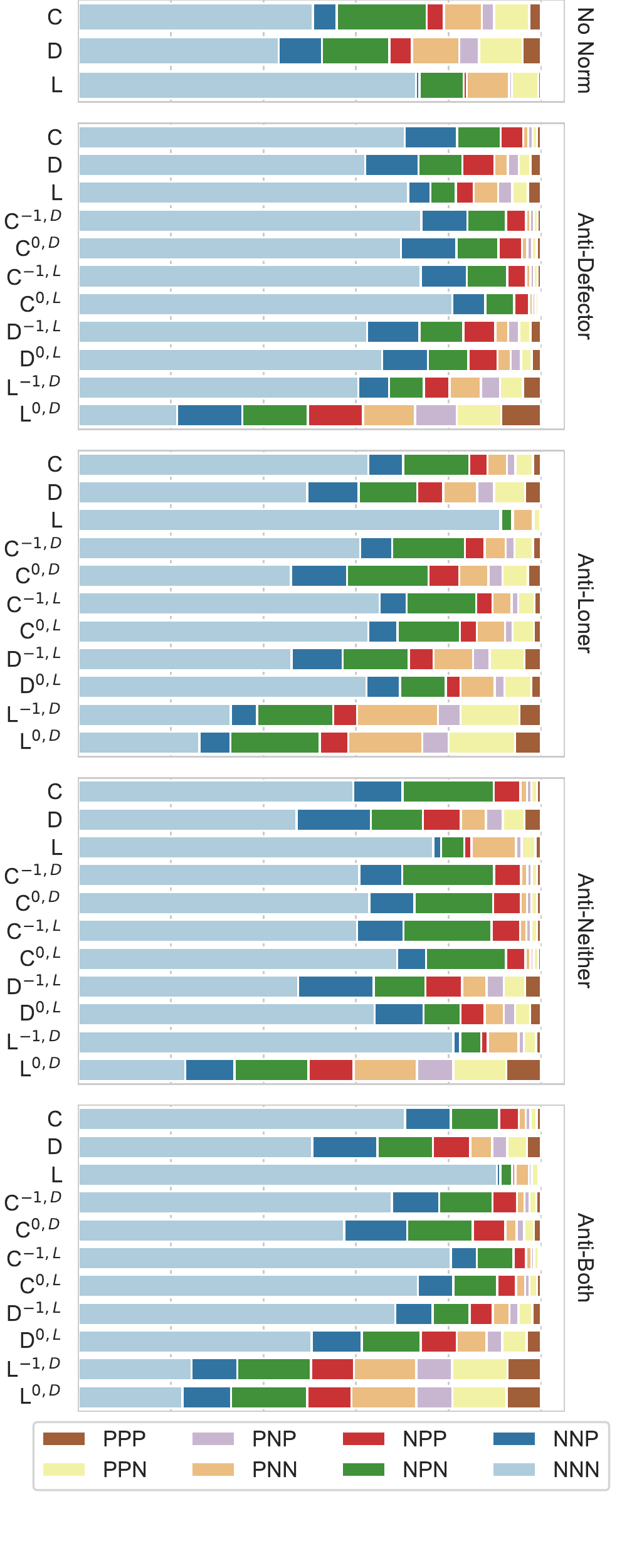}
    \caption{Reputation and Punishment - grouped by strategy and punishment variant}
  \end{subfigure}
  \caption{\textbf{Strategies categorised by punishment-type proportions.} We investigate the proportions of each strategy that utilise a particular punishment variant. In conjunction with our findings, we see that of each strategy that is popular under each social norm, the non-punishing variant is the most popular (see Table \ref{tab:transition_nums}). It should be noted, that the cases where we see larger levels of antisocial punishment (for L\textsupsub{-1,D}{} and L\textsupsub{0,D}{}) are misleading, due to the combination of the strategy's small presence within the population, with the ongoing influx of new punishment variants through mutation. Parameters are identical to those of Fig. \ref{fig:action_distribution} averaged over 100 iterations.}
  \label{fig:punishment_detailed}
\end{figure}

Besides AD, the other social norms excluding AL are also able to increase and stabilise cooperation albeit to a lesser extent (Fig. \ref{fig:action_distribution}). These social norms produce very similar results to the AD social norm both concerning the emerging conditional strategies (Fig. \ref{fig:trans_matrices_condensed_reppun}) and the prevalence of pro-social punishment (SM Figs. \ref{fig:punishment_overview} and \ref{fig:punishment_detailed}). The lower the reputation attributed to loners, the lower the frequency of cooperation and population fitness (Fig. S\ref{fig:payoffs}). Comparing AD and AB against AN social norms (Fig. S\ref{fig:punishment_overview}) shows that as the reputability of being a defector increases, the level of pro-social punishment increases to compensate, thus minimising the damage to cooperation and to the fitness of the population. It follows that populations using comparatively more punishment in addition to reputation have lower average fitness.

Finally, for the same reasons as discussed in the reputation only setup, AL fails to stabilise cooperation altogether, resulting in large fluctuations of cooperation levels, high proportions of the loner strategy in the population, and the lowest payoffs among the social norms both with and without punishment.

\begin{figure}[!h]
  \centering
  \includegraphics[width=\linewidth]{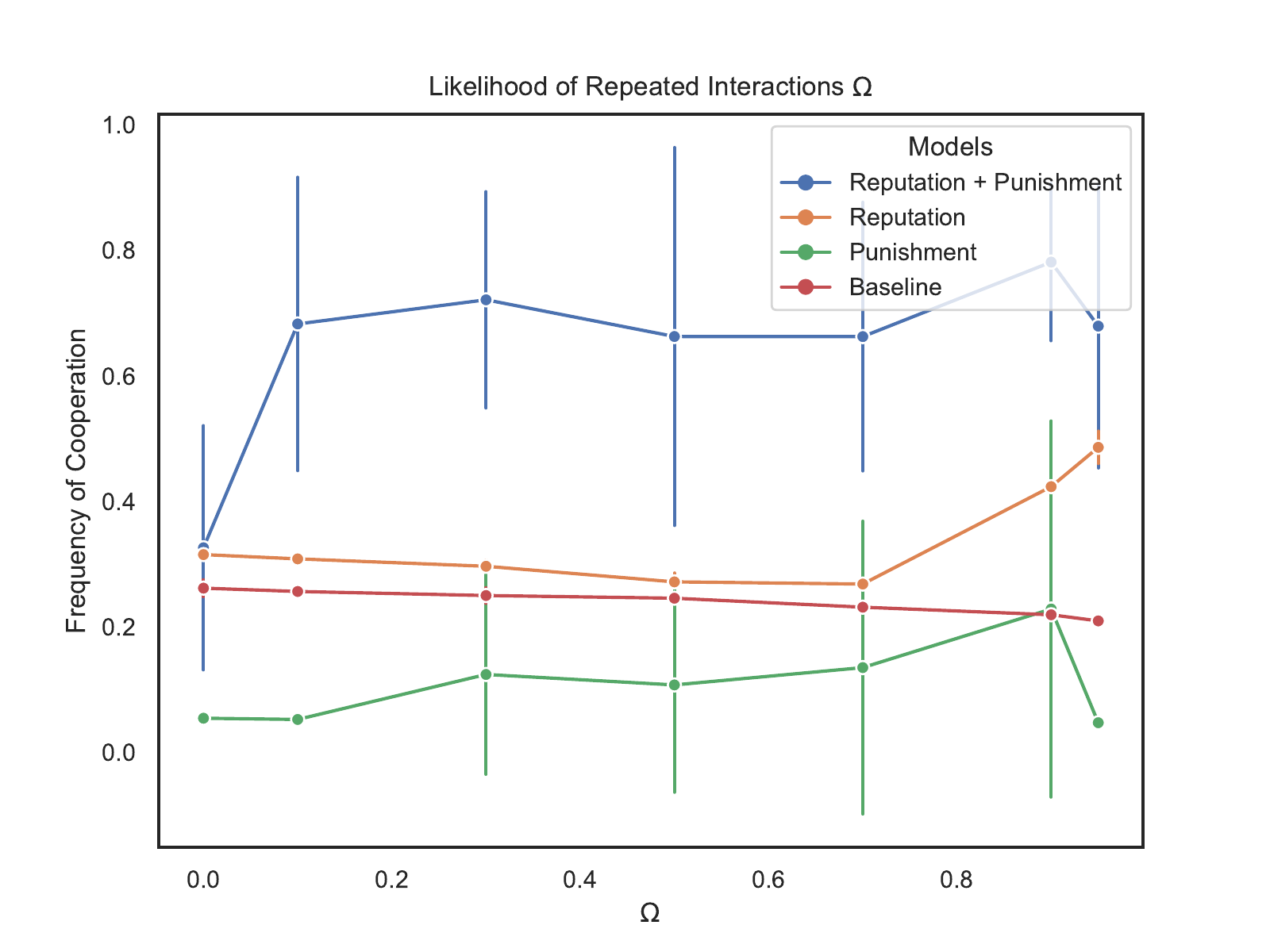}
  \caption{\textbf{The likelihood of further interactions have little impact on the level of cooperation, except in the reputation+punishment model.} Societal systems utilising the whole spectrum of punishment patterns and no reputation) have worse chances for cooperation than systems without any punishment unless $\Omega$ is close to 1. Instead when reputation is present, we see generally identical behaviour to the baseline OPGG except when $\Omega$ is high. Systems using both punishment and reputation generally have stable and high levels of cooperation around 60-80\% when multiple rounds are played each period ($\Omega>0$). However, allowing only a single round each period ($\Omega=0$) results in just over half the former level of cooperation. This shows that both a combination of direct reciprocity (when multiple games are played) and indirect reciprocity (when only a single game is played) is important to obtain high levels of cooperation. Results shown use $N=1000, n=5, r=3, \sigma=1, \gamma=1, \beta=2, t=2 \times 10^5, \epsilon=0.1, m=0.95$ with 10 iterations per data point. }
  \label{fig:omega}
\end{figure}

\begin{figure}[htb]
  \centering
  \includegraphics[width=\textwidth]{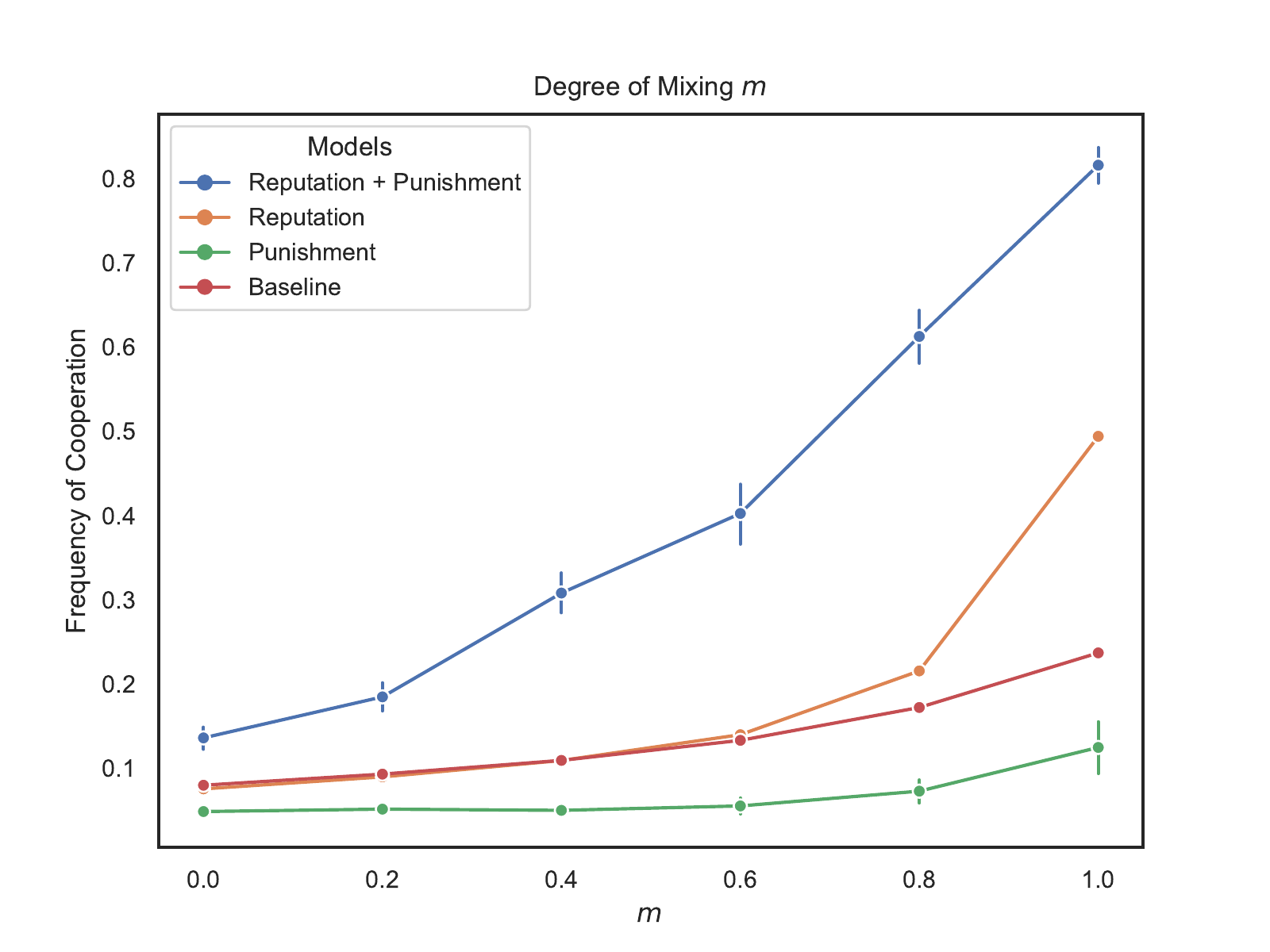}
  \caption{\textbf{More frequent mixing in the evolutionary process leads to higher cooperation.} As the likelihood of learning from another agent from outside the group increases, the frequency of cooperation increases in all cases. Players are more likely to choose an individual with a higher payoff in this larger group to base their strategy evolution upon. Reputation-only populations behave equivalently to the traditional OPGG until $m=0.6$ at which point it shows better chances for cooperation. Populations that implement both the AD social norm and punishment exhibit the same relationship but with generally higher values of cooperation throughout the entire parameter range of $m$. This experiment uses solely the AD social norm, with parameters $N=1000, n=5, r=3, \sigma=1, \gamma=1, \beta=2, t=2 \times 10^5, m=0.95, \Omega=\tfrac{10}{11}$ with 10 iterations per data point.}
  \label{fig:m_actions}
\end{figure}

\paragraph{Parameter robustness}
Our main results discussed so far are robust to an extensive set of parameter variations. These are reported in the Supplementary Materials, Figs. S\ref{fig:r_actions} - S\ref{fig:epsilon_actions} where the effects of changes to the group synergy factor $r$, the loner's payoff $\sigma$, the punishment cost $\gamma$ and penalty $\beta$, probability of repeated interactions $\Omega$, OPGG group size $n$, the degree of evolutionary group mixing $m$, and the rate of mutation $\epsilon$ are systematically explored (the parameterisations used for each robustness check are reported in SM Table S\ref{parameters}). The superiority of combining the reputation and punishment mechanisms remain consistent with our findings, showing higher levels of cooperation when compared to our baseline model without reputation or punishment, and to the baseline model with only punishment. While the model utilising solely reputation does tend to increase cooperation, its behaviour is somewhat more dependent on specific parameter ranges of the degree of mixing $m$ (SM Fig. S\ref{fig:m_actions}) and of the stability of groups $\Omega$ (SM Fig. S\ref{fig:omega}). Cooperation increases when the degree of mixing $m$ increases, likely due to the pool of available strategies for agents to choose to evolve to. When $m$ is small, players evolve based on a randomly selected player within their own OPGG groups. The likelihood of a fruitful and cooperative player to exist within this group is lower than when $m$ is high and players evolve probabilistically based on the payoffs of a randomly selected individual from outside of their group. Instead, considering the likelyhood of further interactions $\Omega$, in the setup in which both reputation and punishment are active, as long as $\Omega \neq 0$, cooperation has a high and similar proportion within the population. This suggests that a combination of both direct (relying on direct experience within the group to condition action) and indirect reciprocity (rely on third party information to condition action) is important in obtaining high levels of cooperation.

\section{Discussion}

The OPGG provides a challenging environment for cooperation. The three-way interaction between cooperators, defectors and loners produce natural cyclical dynamics, where cooperative environments favour defection, which in turn makes opting out more advantageous, which in turn leads to the return of cooperation \citep{Hauert2002,Hauert2002a}. Similar challenges are faced introducing the possibility of agents punishing one another based on their past actions. In this case the cyclical dynamics involve both the presence of pro-social (punishment of defectors and loners) and of anti-social punishment (punishment of cooperators), with each strategy punishing its would-be invader \citep{Rand2011}. In either case, the average cooperation level resulting from the cyclical behaviour is low.

Introducing into the OPGG the AD social norm (assigning cooperators a good reputation and assigning better reputations to agents that opt-out from the game than to those who participate but withhold their contribution), significantly increases cooperation levels unlike the other reputational systems. The effectiveness of the AD norm depends on the fact that it provides a way for agents to condition their behaviour to the propensity of their group to cooperate, while reducing the attractiveness of the loner action, which remains as a backup if the environment becomes too hostile. Accordingly, within the ecosystem of strategies generated by the introduction of this norm, over half of the population end up preferring one of two strategies, C\textsupsub{0,L}{} or D\textsupsub{0,L}{}. Both act as loners if the environment becomes too harsh, but turn to cooperation or defection in a more friendly world. The backup option of abstaining from the OPGG on the one hand prevents defectors from over-exploiting cooperators: if they are present in sufficient numbers, their bad reputation induces C\textsupsub{0,L}{} to become a loner. On the other hand, the same rule supports a conditional defector who is even evolutionarily stronger than the unconditional counterpart, as it viciously exploits reputation and cooperative behaviour while still receiving the loner's payoff when the likelihood of exploiting its group decreases. The evolutionary success and average payoffs of these two strategies are similar, hence agents cycle between them, with their equilibrium favouring conditional cooperators, hence the observed increase in cooperation. It should be noted that in this setup, many other strategies survive - albeit in lower numbers - thus the dominance of the aforementioned strategies is never complete. It is interesting to note that the laxer counterparts of the strategies discussed, namely C\textsupsub{-1,L}{} and D\textsupsub{-1,L}{} are less successful, as they propose cooperation or defection even in groups largely dominated by defectors.

\begin{figure}[!h]
  \centering
  \includegraphics[width=\textwidth]{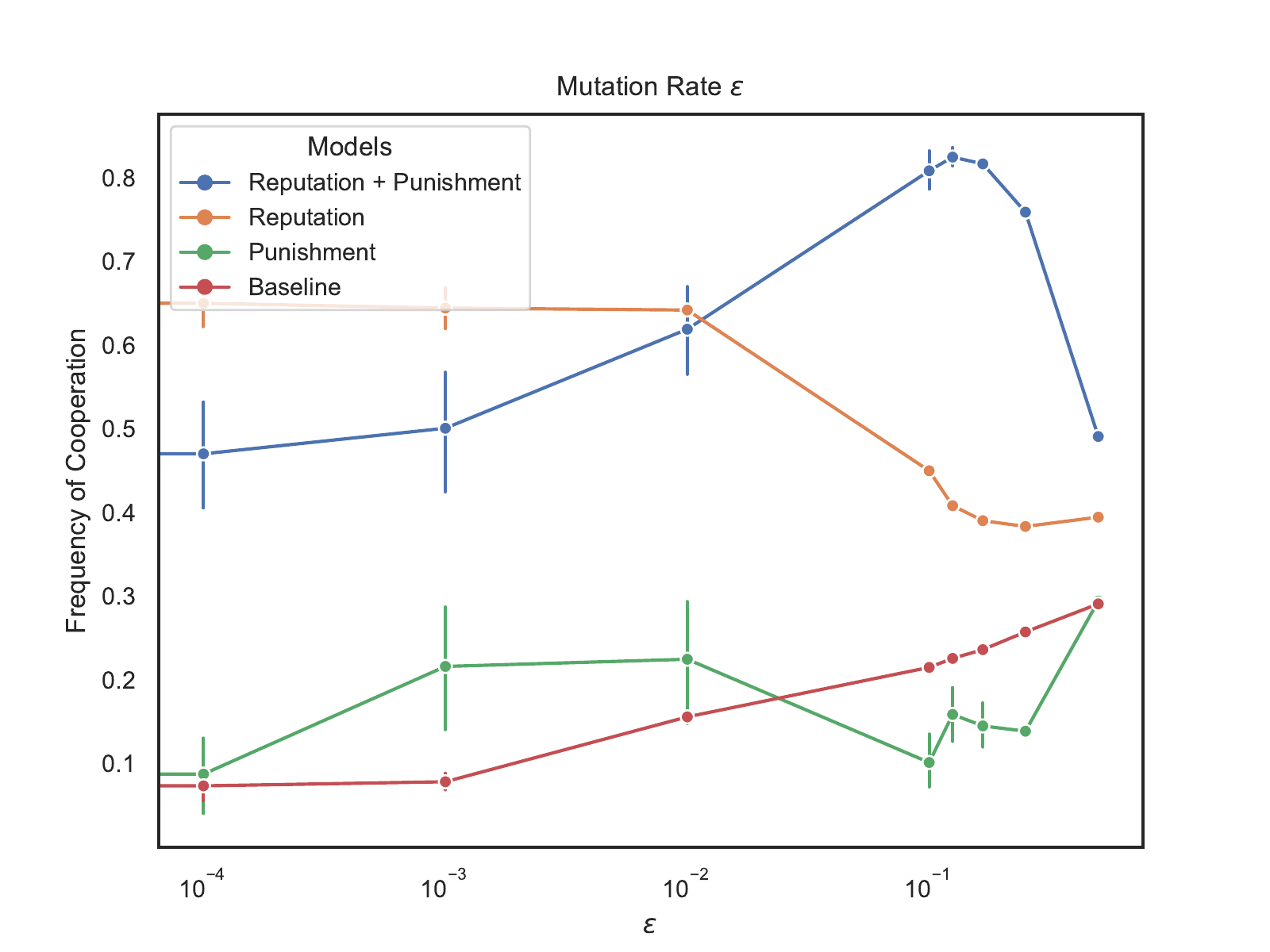}
  \caption{\textbf{Higher mutation rates required for cooperation with reputation and punishment, but lower rates preferred with only reputation.} Populations that utilise both a social norm and punishment mechanism, achieve the highest levels of cooperation when mutants are introduced often, between 1-4 mutants per 10 time-periods. When mutants are added more frequently, we see a sudden drop in cooperation. This is due to non-cooperative strategies being inserted into the population faster than evolution can turn them into cooperative players. The same range of mutation rates $\epsilon$ result in the worst levels of cooperation for populations that use only a social norm and not punishment. In summary, reputation-only setups fare best with lower mutation rates, while populations where punishment is also available result in high levels of cooperation also for high mutation rates. This experiment uses solely the AD social norm, with parameters $N=1000, n=5, r=3, \sigma=1, \gamma=1, \beta=2, t=2 \times 10^5, m=0.95, \Omega=\tfrac{10}{11}$ with 25 iterations per data point.}
  \label{fig:epsilon_actions}
\end{figure}

Cooperation turns out to be much more likely when we include the interaction between costly punishment and a reputational system into the mix, assigning a better reputation to agents that opt-out from the game than to those that participate but withhold their contribution. It should be noted that the emergence of high levels of cooperation in this setup is surprising as the presence of anti-social punishment has been found to reduce the levels of cooperation (\citealp{Rand2011} and replicated in Fig. \ref{fig:action_distribution}) in the OPGG. Given that the success of the combination of punishment and reputation in sustaining cooperation is much higher than the success of either of the two alone, the two mechanisms act in synergy with each other. The ecosystem of strategies that thrive in this environment is made up of strategies whose main (or sole) action is cooperation, who for the most part do not employ the punishment device at all, with a smaller but material proportion punishing pro-socially.
For each conditional strategy, there are significant mutual transitions between the non-punishing and the pro-socially punishing variants, indicating a minor but present role for the latter in preserving cooperation. This cooperative environment relies on the regular influx of mutant strategists to mitigate the second-order free-rider problem, moderating the cyclical dynamics it induces. New punishing agents introduced into the population are progressively transformed into the non punishing variants. Their role is confirmed by SM Fig. S\ref{fig:epsilon_actions} where in the presence of both reputation and punishment, cooperation levels decrease for lower mutation rates.

All in all, these results point to the fact that in the OPGG, reputational information acts as a cheap substitute to costly punishment. Cooperators are rewarded with a positive reputation, which makes future opponents more likely to cooperate with them. This is in line with the experimental results of \cite{Ule2009}, who found that in the presence of reputation, subjects prefer rewarding positive behaviour to punishing those who defect.Since the AD reputational mechanism can be considered as an indirect method of sanctioning anti-social behaviour, our result is also in line with \cite{Balafoutas2014} which experimentally finds that people prefer to punish indirectly rather than directly.

Introducing a reputational system has a differential impact on the different types of punishment.
In the absence of reputation, anti-social punishment is able to successfully prevent the invasion of defectors by cooperators -- as in \cite{Rand2011} -- but is unable to do the same with the new conditional strategies under a social norm. Contrastingly, pro-social punishment positively reinforces the reputational information, increasing the level of cooperation in the population. This result is in contrast to \cite{Gurerk2006}, that establishes an experimental superiority of punishment institutions over reputational mechanisms for the public goods game and poses doubts on whether this result can be extended to experiments where the OPGG is played.

It should be noted that all strategies with cooperation as the main action are selected in the ecosystem for this setup, and in remarkably similar proportions. This follows from the fact that they always act as cooperators in a strongly cooperative environment, thus becoming indistinguishable from each other (thereby avoiding selection).

Our results qualify the findings of the experimental literature on anti-social punishment \citep{Herrmann2008,Gachter2009,Gachter2011}. Such contributions work with designs that rigidly separate individual choices about actions and punishments from reputational concerns. This is particularly true for \cite{Rand2011} which was performed on Amazon Mechanical Turk, where professionals play many games. In recent experiments, run in-person and with samples of the general population where individuals knew they were playing with citizens from the same villages or small towns \citep{Pancotto2020}, anti-social punishment was not significant. In these cases, even within an experimental setup, social norms  implicitly matter and remove the scope for anti-social punishment.

Reproducing in-silico a simplified version of the combination of social reputation mechanism and punishment systems that characterise complex societies, our contribution shows that these two dimensions interact, shedding light on the surprising success of reputation in a world under the contemporaneous threat of exploitation and of anti-social punishment. Finally, our results contribute to identifying the conditions that allow effective collective action in the presence of the possibility to opt-out of interactions.

\section*{Acknowledgements}
Simone Righi gratefully acknowledges funding from ``Decentralised Platform Economics: Optimal Incentives and Structure'', Fondi primo insediamento, Ca'Foscari University of Venice.
Francesca Pancotto gratefully acknowledges funding from ``Data driven methodologies to study social capital and its role for economic growth'', University of Modena and Reggio Emilia, FAR 2019, CUP E84E19001120005.
Shirsendu Podder gratefully acknowledges doctoral funding from the Engineering and Physical Sciences Research Council. The authors acknowledge the use of the UCL Myriad High Performance Computing Facility (Myriad@UCL), and associated support services, in the completion of this work.

\bibliography{Paper}


\newpage
\date{}
\title{Reputation and Punishment sustain cooperation in the Optional Public Goods Game Supplementary Information}
\maketitle
\begin{abstract}
  Cooperative behaviour has been extensively studied as a choice between cooperation and defection. However, the possibility to not participate is also frequently available. This type of problem can be studied through the optional public goods game. The introduction of the ``Loner'' strategy, allows players to withdraw from the game, which leads to a cooperator-defector-loner cycle. While prosocial punishment can help increase cooperation, anti-social punishment -- where defectors punish cooperators -- causes its downfall in both experimental and theoretical studies.

  In this paper, we introduce social norms that allow agents to condition their behaviour to the reputation of their peers. We benchmark this both with respect to the standard optional public goods game and to the variant where all types of punishment are allowed. We find that a social norm imposing a more moderate reputational penalty for opting out than for defecting, increases cooperation. When, besides reputation, punishment is also possible, the two mechanisms work synergically under all social norms that do not assign to loners a strictly worse reputation than to defectors. Under this latter setup, the high levels of cooperation are sustained by conditional strategies, which largely reduce the use of pro-social punishment and almost completely eliminate anti-social punishment.
\end{abstract}
\keywords{Reputation, anti-social punishment,  optional public goods game}

\pagebreak


\section{Social Norm Alternatives}

\subsection{Anti-Defector Transition Matrices}
\begin{figure}[h]
  \centering
  \includegraphics[width=0.79\linewidth]{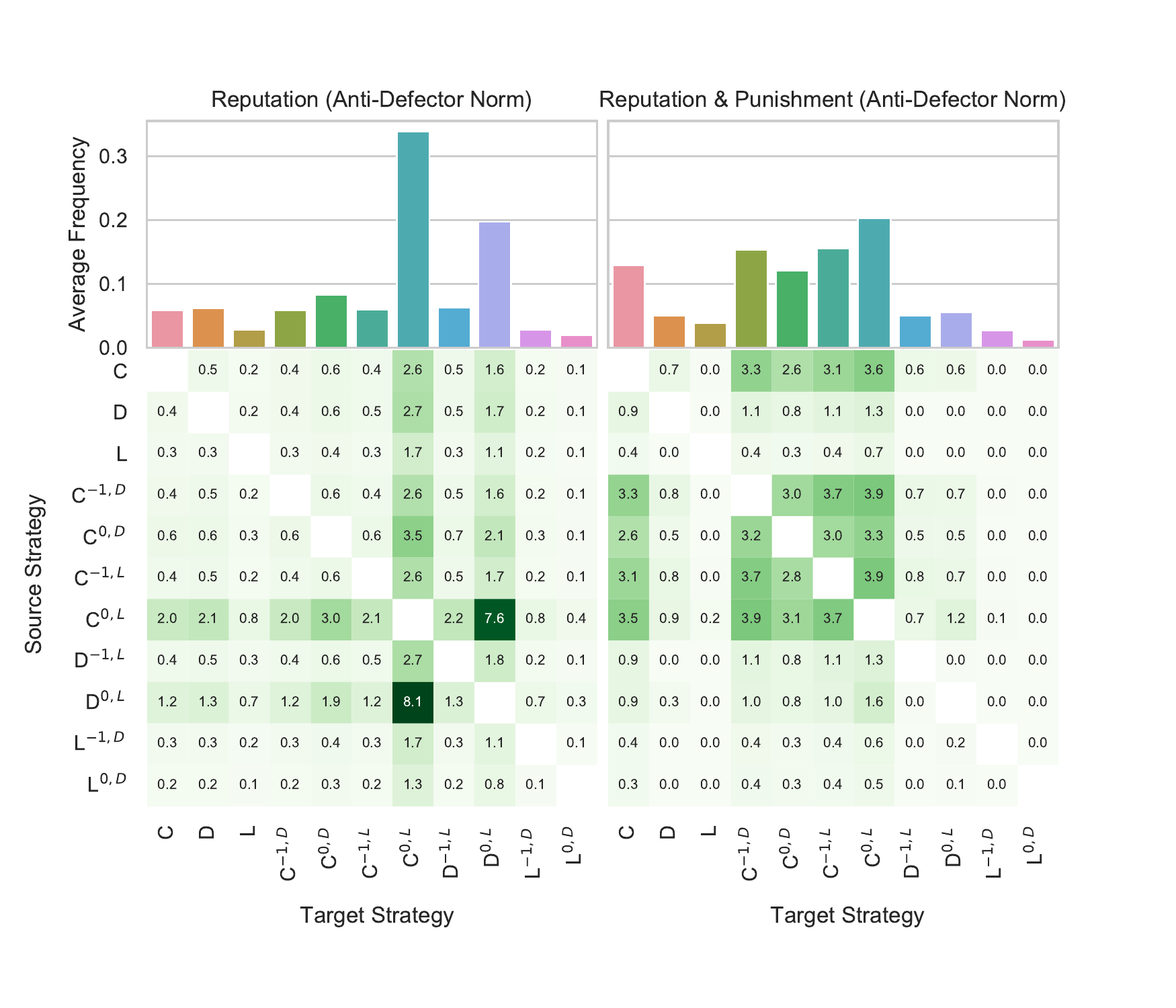}
  \caption{\textbf{Transition matrices with the  AD social norm, with (Right Panel) and without (Left Panel) punishment.} As per the transition networks in Figs. \ref{fig:trans_matrices_condensed_rep} and \ref{fig:trans_matrices_condensed_reppun} in the main text, in the case with only reputation, the most popular strategies are C\textsupsub{0,L}{} and D\textsupsub{0,L}{}, while in the case also with punishment, the most popular strategies are those with cooperation as their main action in very similar levels to each other. The Reputation \& Punishment matrix has been condensed by summing over the punishment variants for each of the 11 strategies. Values shown represent the proportion of times a player from a given strategy moved to another (regardless of punishment variant), relative to the total number of transitions across the population. Simulation parameters are identical to those in Fig. \ref{fig:action_distribution} with 100 iterations per parameterisation. Numerical values in the matrices represent percentages of the total number of transitions in the entire population, excluding all transitions within the same strategy but to a different punishment variant.}
  \label{fig:transition_matrix_AD}
\end{figure}
\pagebreak

\subsection{Anti-Loner Transition Matrices}
\begin{figure}[h]
  \centering
  \includegraphics[width=0.79\linewidth]{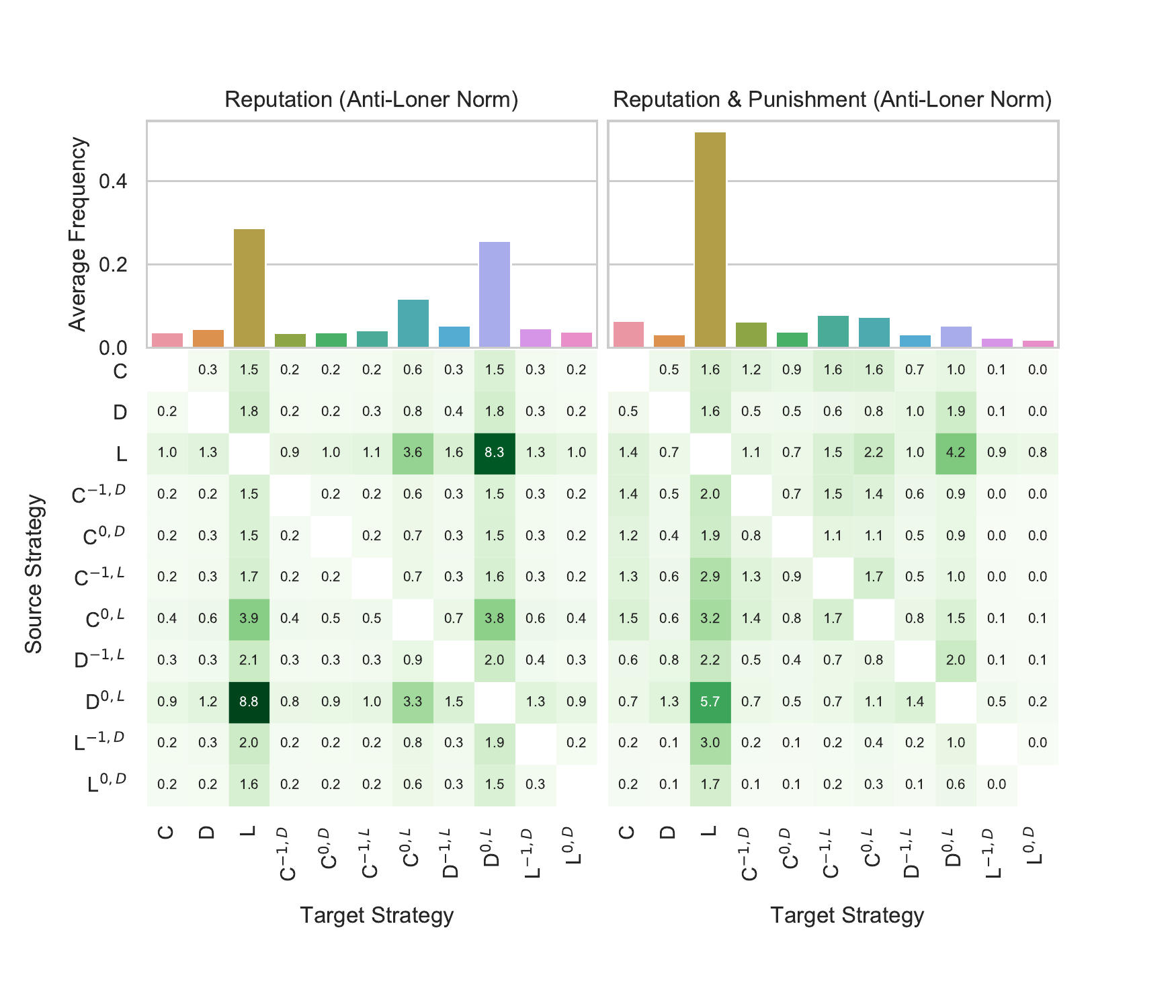}
  \caption{\textbf{Transition matrices with the  AL social norm, with (Right Panel) and without (Left Panel) punishment.} In both models, populations become composed mostly of loners. Under reputation and punishment, the loners behave unconditionally, whereas under reputation, loners act under a combination of L and D\textsupsub{0,L}{} strategies, and C\textsupsub{0,L}{} to a lesser extent. Simulation parameters are identical to those in Fig. \ref{fig:action_distribution} with 100 iterations per parameterisation. Numerical values in the matrices represent percentages of the total number of transitions in the entire population, excluding all transitions within the same strategy but to a different punishment variant.}
  \label{fig:transition_matrix_AL}
\end{figure}
\pagebreak

\subsection{Anti-Neither Transition Matrices}
\begin{figure}[h]
  \centering
  \includegraphics[width=0.79\linewidth]{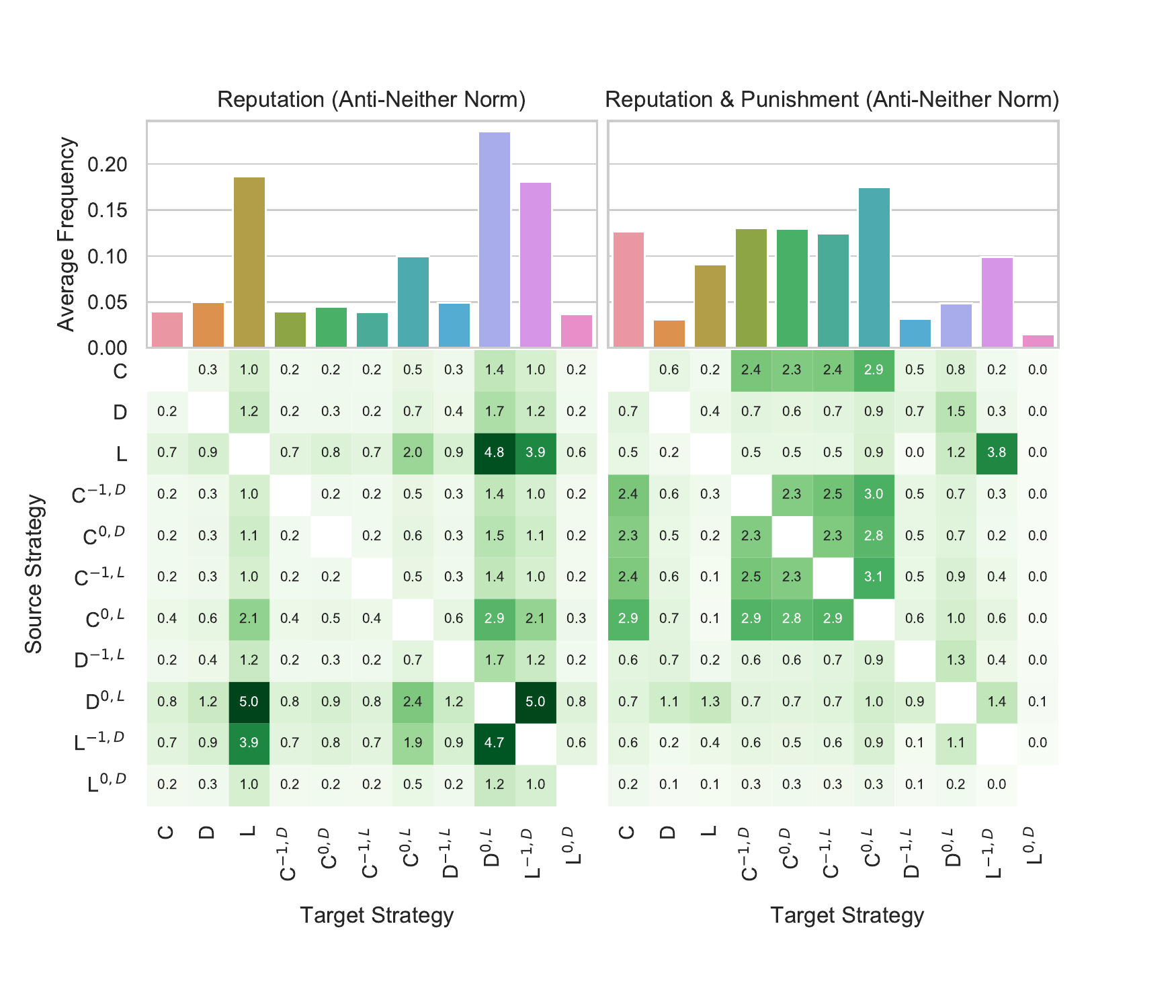}
  \caption{\textbf{Transition matrices with the  AN social norm, with (Right Panel) and without (Left Panel) punishment.} The dynamics of this social norm are very similar to those of the AD norm, with the notable exception of the loner strategies under both models. Assigning a better reputation to defectors causes a rise in non-cooperative behaviour. This in turn forces conditional strategies to utilise their backup option of abstaining from the game in order to protect themselves from exploitation, leading to the higher levels of (conditional and unconditional) loner strategists in the population. Simulation parameters are identical to those in Fig. \ref{fig:action_distribution} with 100 iterations per parameterisation. Numerical values in the matrices represent percentages of the total number of transitions in the entire population, excluding all transitions within the same strategy but to a different punishment variant.}
  \label{fig:transition_matrix_AN}
\end{figure}
\pagebreak

\subsection{Anti-Both Transition Matrices}
\pagebreak

\subsection{Intra-strategy Transition Matrices}

\begin{figure}[h!]
  \centering
  \includegraphics[scale=0.79]{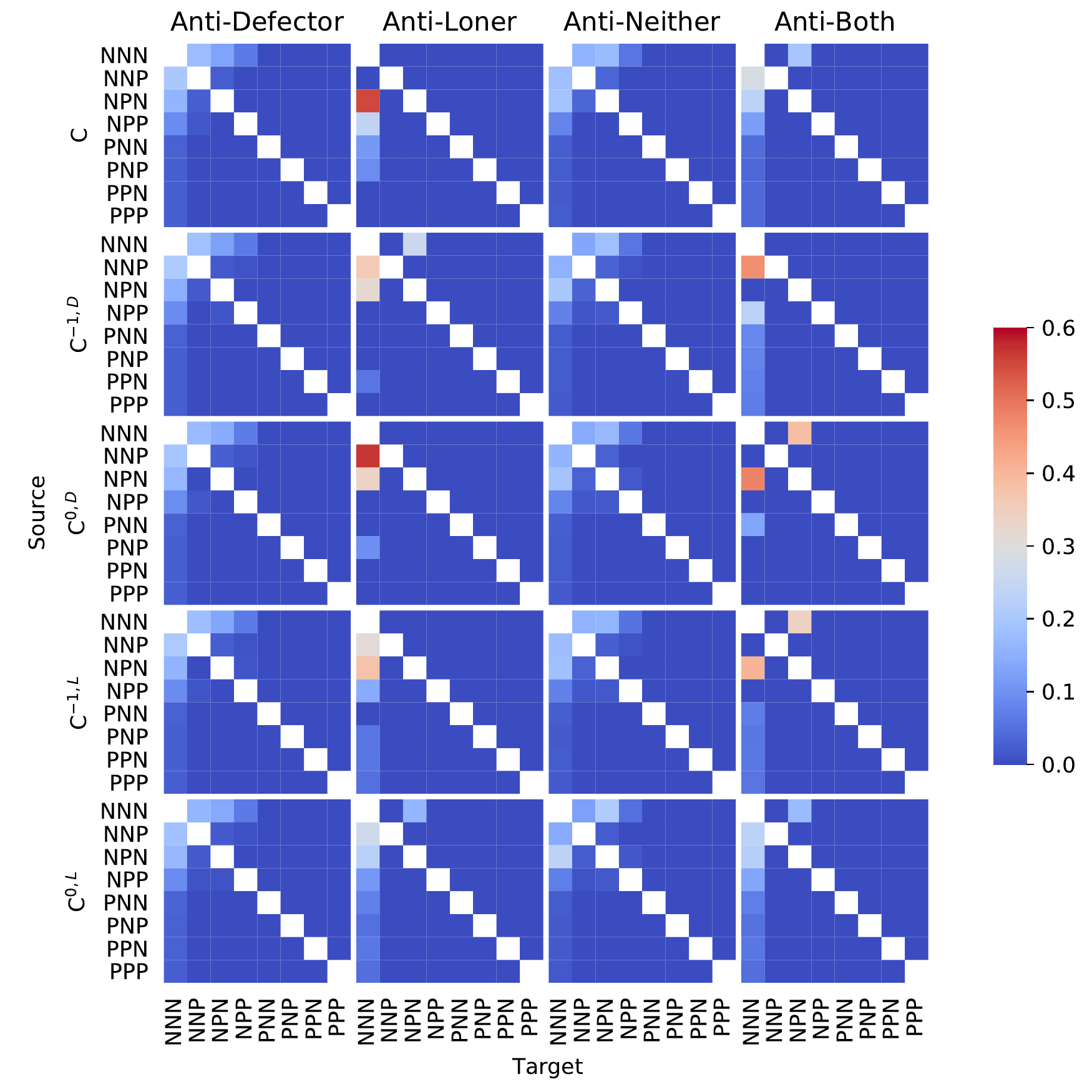}
  \caption{\textbf{Transition matrices between punishment variants of cooperative strategies.} Ignoring the transitions between strategies (e.g. from C\textsupsub{-1,D}{} to C\textsupsub{0,L}{}), here we focus on the transitions between punishment variants of the same strategy (e.g. from C\textsupsub{-1,D}{NPP} to C\textsupsub{-1,D}{NNN}) for populations utilising both reputation and punishment. The four columns represent the four possible social norms, and the five rows represent the five strategies that cooperate as their primary action.
    Within each heat-map, the strongest transitions occur either in the first column (for the transitions towards the non-punishing variant) or the first row (for the transitions away from the non-punishing variant). There are no exceptions to this for the generally cooperative strategies that compose the majority of the population under the Reputation + Punishment model. Of the transitions away from the non-punishing variant, the most popular targets tend to be those that punish pro-socially, punishing defectors, loners or both. There is a clear absence of strategies that punish anti-socially. Simulation parameters are identical to those in Fig. \ref{fig:action_distribution} with 100 iterations per parameterisation. The colour intensity represents the movement between punishment variants, relative to the total number of transitions between variants of that strategy.}
  \label{fig:intrastrategy_rep_pun_transitionmatrix_C}
\end{figure}
\pagebreak
\begin{figure}[h!]
  \centering
  \includegraphics[scale=0.79]{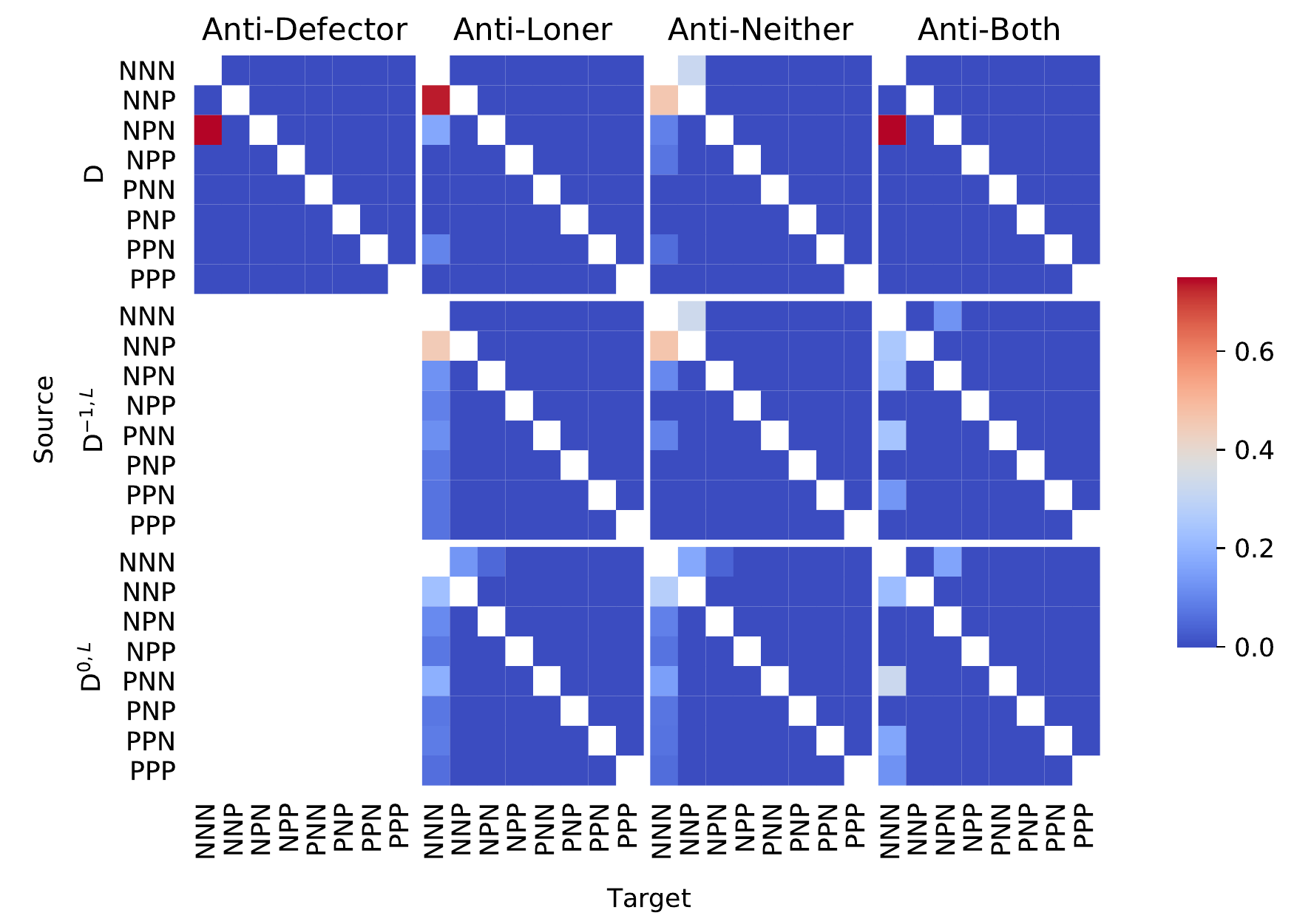}
  \caption{\textbf{Transition matrices between punishment variants of non-cooperative strategies.} Generally non-cooperative strategies exhibit very similar behaviour to those that primarily cooperate. Under the Anti-Defector norm, the D\textsupsub{-1,L}{} and D\textsupsub{0,L}{} strategies do not present any transitions to a punishment variant of the same strategy, most likely due to the inherent weakness of the strategy within the population. The largest transitions in this subset of strategies tend to be players moving away from the punishment variants that punish defectors and/or loners. Simulation parameters are identical to those in Fig. \ref{fig:action_distribution} with 100 iterations per parameterisation. The colour intensity represents the movement between punishment variants, relative to the total number of transitions between variants of that strategy.}
  \label{fig:intrastrategy_rep_pun_transitionmatrix_D}
\end{figure}
\pagebreak
\pagebreak


\subsection{Population Composition}

\pagebreak

\subsection{Population Composition Alternative}

\pagebreak

\subsection{Payoffs}
\pagebreak

\subsection{Baseline OPGG - with and without Punishment}
\begin{figure}[h]
  \centering
  \includegraphics[width=0.6\linewidth]{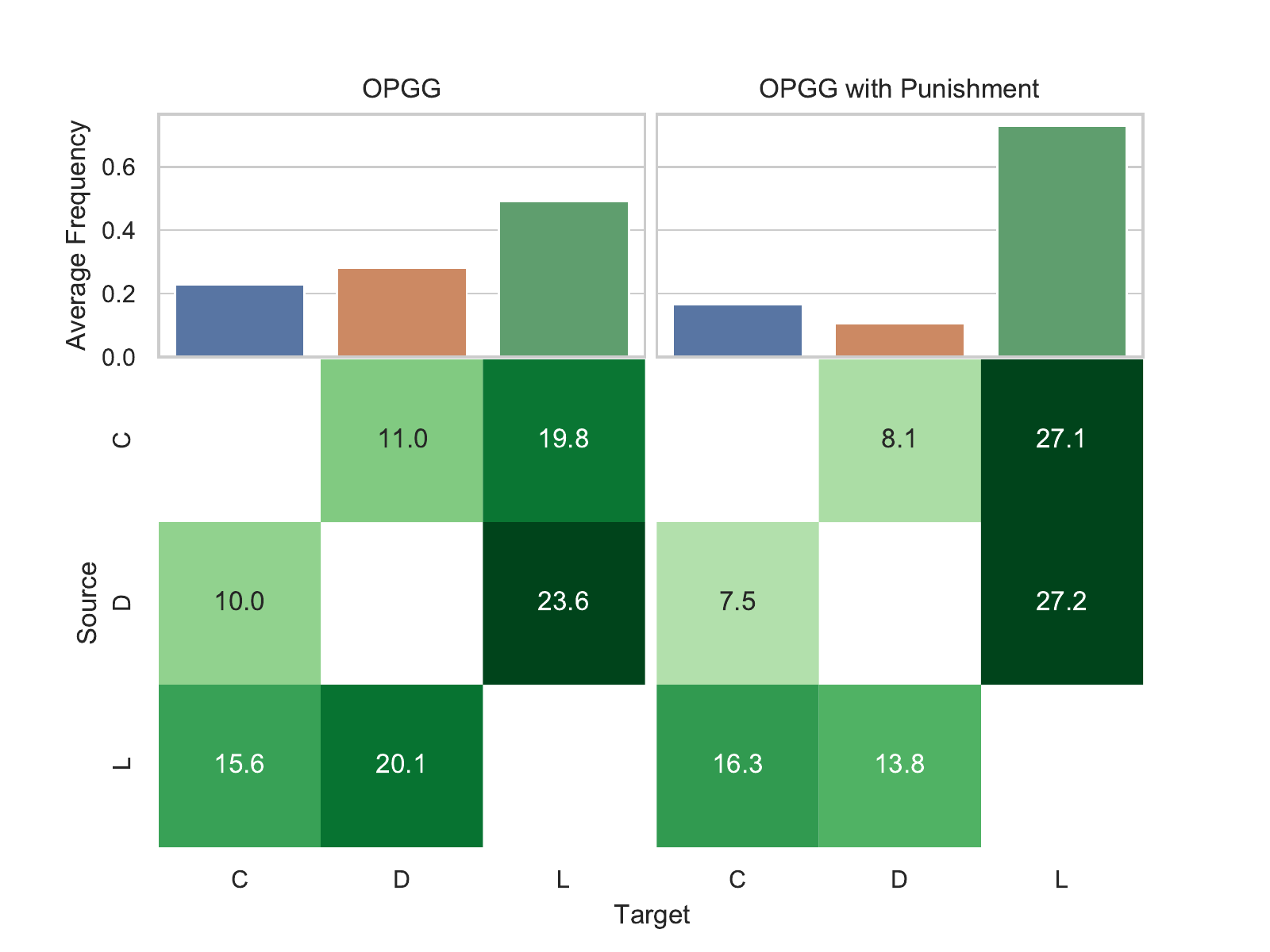}
  \caption{\textbf{OPGG Transition matrices and average frequencies of baseline strategies for baseline models.} Here we reproduce the transition matrices for the traditional OPGG with no punishment or reputation, and the OPGG with all punishment strategies available. What we observe in the transition matrices and analysing the time series fluctuations in the population state in Panel A of Fig. \ref{fig:series} is clearly cyclic behaviour. In the absence of reputation and punishment, we see only around 20\% cooperation in comparison to the model with punishment, where we see a loner dominated population, with only around 12\% cooperation.}
  \label{fig:transition_matrix_baseline}
\end{figure}


\pagebreak
\section{Punishment Patterns}
\subsection{Overview}

\pagebreak
\subsection{Detailed Breakdown}

\pagebreak

\subsection{Transitions to and from strategies with and without Punishment}

\begin{table}[h]
  \centering
  \begin{tabular}{lllll}
    \toprule
                                   &                                         &                        & \multicolumn{2}{c}{\textbf{To}}                          \\ \toprule
                                   &                                         &                        & \textit{Non-Punishing}          & \textit{Punishing}     \\
    \multirow{8}{*}{\textbf{From}} & \multicolumn{1}{c}{\multirow{2}{*}{AD}} & \textit{Non-Punishing} & 113945 (47.9\%)                 & 39562 (16.6\%)         \\
                                   & \multicolumn{1}{c}{}                    & \textit{Punishing}     & 74846 \hfill (31.4\%)           & 9714 \hfill  (4.1\%)   \\ \cline{2-5}
                                   & \multirow{2}{*}{AL}                     & \textit{Non-Punishing} & 80875 \hfill (53.2\%)           & 12033 \hfill  (7.9\%)  \\
                                   &                                         & \textit{Punishing}     & 58639 \hfill (35.6\%)           & 607 \hfill  (0.4\%)    \\ \cline{2-5}
                                   & \multirow{2}{*}{AN}                     & \textit{Non-Punishing} & 93443 \hfill (45.2\%)           & 34749 \hfill  (16.8\%) \\
                                   &                                         & \textit{Punishing}     & 66390 \hfill (32.1\%)           & 12315 \hfill  (6.0\%)  \\ \cline{2-5}
                                   & \multirow{2}{*}{AB}                     & \textit{Non-Punishing} & 104659 \hfill (67.1\%)          & 9008 \hfill  (5.8\%)   \\
                                   &                                         & \textit{Punishing}     & 42109 \hfill (27.0\%)           & 240 \hfill  (0.2\%)    \\ \bottomrule
  \end{tabular}
  \caption{\textbf{Overall, punishment is not popular. }In line with our results, the majority of transitions in each social norm are either between non-punishing strategies, or are from a punishing strategy moving to a non-punishing strategy. We note that regardless of the social norm, the proportion of players transitioning from a non-punishing strategy to a punishing-strategy is always at least half of the proportion of players moving from a punishing strategy to a non-punishing strategy. All values are the number of transitions (rounded to the nearest integer) between punishing and non-punishing strategies averaged over 100 simulations. Parameters are identical to those of Fig. \ref{fig:action_distribution}.}
  \label{tab:transition_nums}
\end{table}


\pagebreak
\section{Illustrative Time Series}
\label{sm:timeseries}

\subsection{Anti-Defector Social Norm}

\begin{figure}[h]
  \centering
  \includegraphics[width=0.65\linewidth, trim=0 5 0 23, clip]{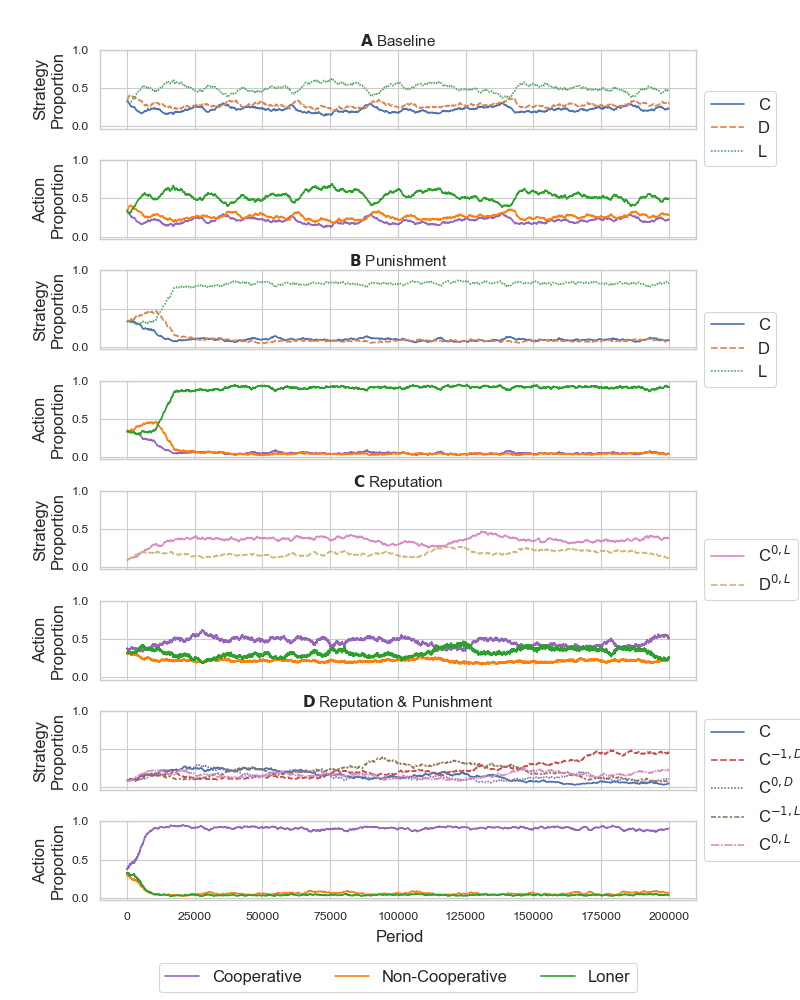}
  \caption{\textbf{Time series representations of singular simulations.} Proportion of strategies (top)  and actions (bottom) within the population for each of the four central models. The Baseline with no punishment or reputation (panel A), the Baseline with only punishment (B), the model with only reputation (C), and the model with reputation and punishment (D). Only the relevant subset of strategies that remain in notable amounts in the population. Panel A shows evidence of cyclic behaviour between the three strategies. Panel B shows the rapid failure of cooperators and the inevitable domination of loners. Panel C shows the equilibrium reached between C\textsupsub{0,L}{} and D\textsupsub{0,L}{} with reasonable and generally stable levels of cooperation. Finally, Panel D shows the prevalence of the cooperative strategies, as well as very stable levels of cooperation in the population.}
  \label{fig:series}
\end{figure}

\pagebreak
\subsection{Reputation only}

\pagebreak
\subsection{Reputation \& Punishment}

\begin{figure}[h]
  \centering
  \includegraphics[width=0.65\linewidth, trim=0 5 0 23, clip]{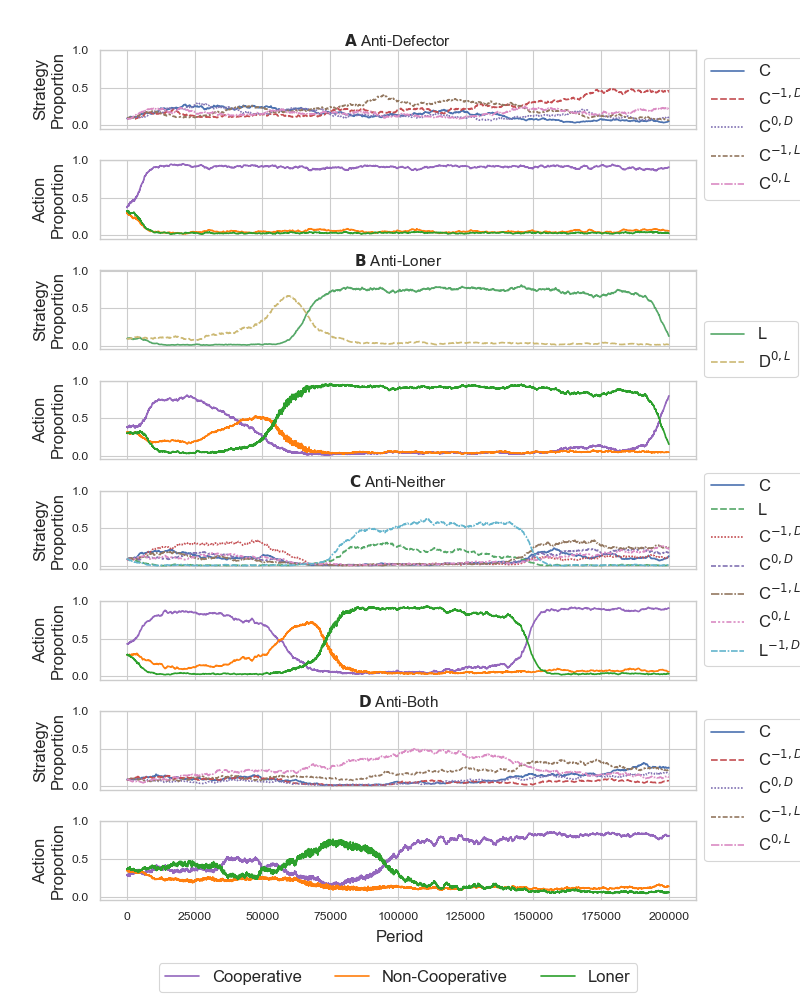}
  \caption{\textbf{Time series representations of singular simulations, reputation and punishment.} The most interesting dynamics emerge when reputation works in conjunction with punishment. Panel A shows a very stable level of cooperation with a largely cooperative population. Panels B-D show cyclic behaviour between cooperators, defectors and loners. In Panels B and D, the AL and AB social norms both attribute the worst reputation to being a loner, however in the AB social norm, defection is seen to be an equally bad action. This reinforces our finding that as long as being a loner is not worse than being a defector, it remains an option for cooperators looking for a safe haven from exploitation. }
  \label{fig:series_reppun}
\end{figure}


\pagebreak
\section{Parameter Explorations}

In this section, data points and error bars represent the averages and standard deviations of the overall level of cooperation over several runs with identical parameter sets (see Table \ref{parameters}). The overall level of cooperation in a single simulation is measured by the average of cooperative actions in the latter half of the simulation. These are then averaged over several simulations.

In the following, we focus on the Anti-Defector social norm where defectors are attributed a reputation of -1 and loners are attributed a reputation of 0. As always, cooperators are attributed a reputation of 1.

\subsection{Group synergy factor}

\begin{figure}[h]
  \centering
  \includegraphics[width=\linewidth]{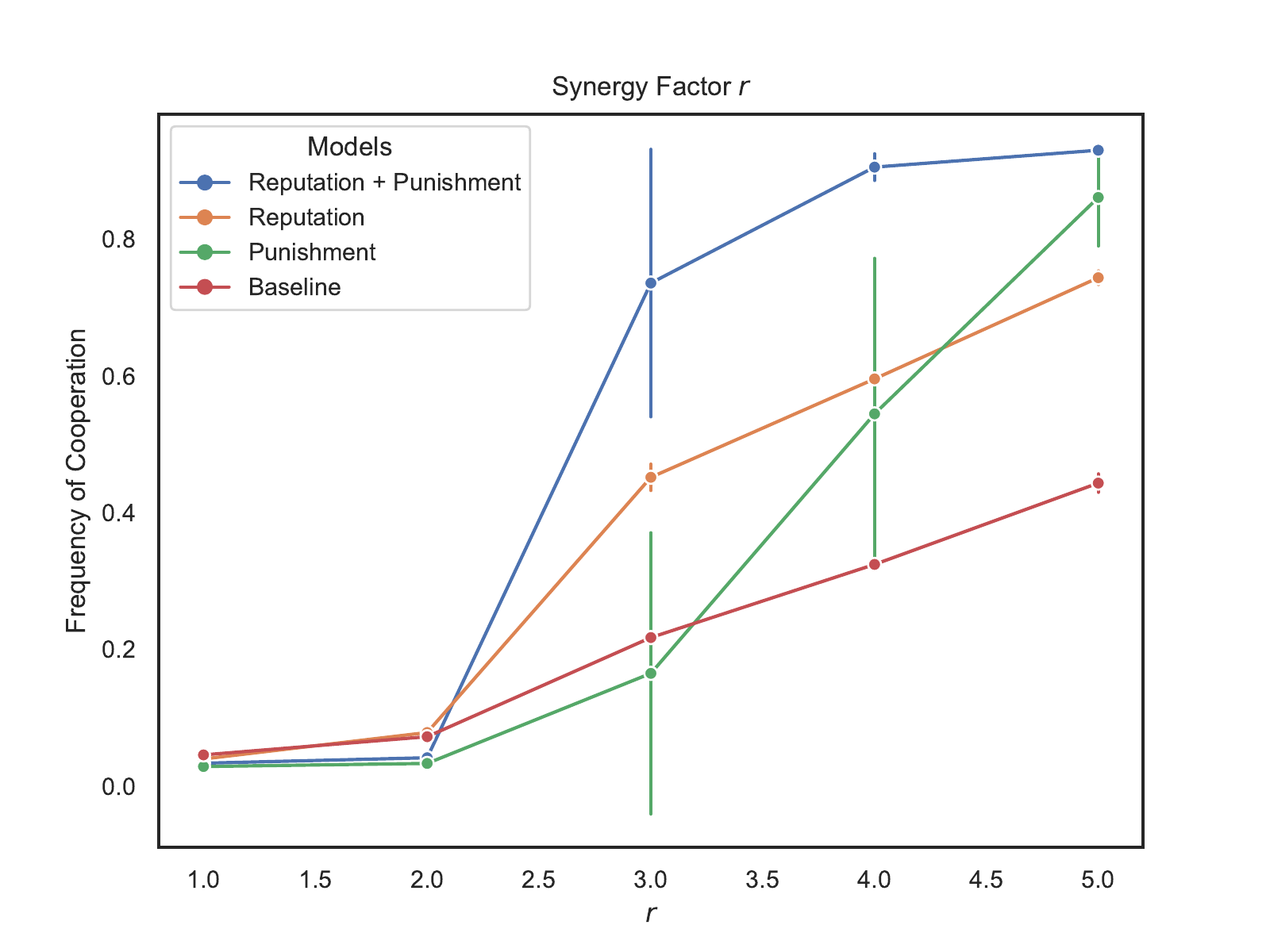}
  \caption{\textbf{Cooperation increases when the OPGG multiplier increases.} In line with the findings of the traditional OPGG, cooperation increases as $r$ increases when both reputation is used, and when both reputation and punishment is used. The latter model results in the greatest increase. Results shown use $N=1000, n=5, \sigma=1, \gamma=1, \beta=2, t=2 \times 10^5, \epsilon=0.1, m=0.95$ and $\Omega=\tfrac{10}{11}$ with 25 iterations per data point.}
  \label{fig:r_actions}
\end{figure}

\pagebreak
\subsection{Loner's Payoff}

\begin{figure}[h]
  \centering
  \includegraphics[width=\linewidth]{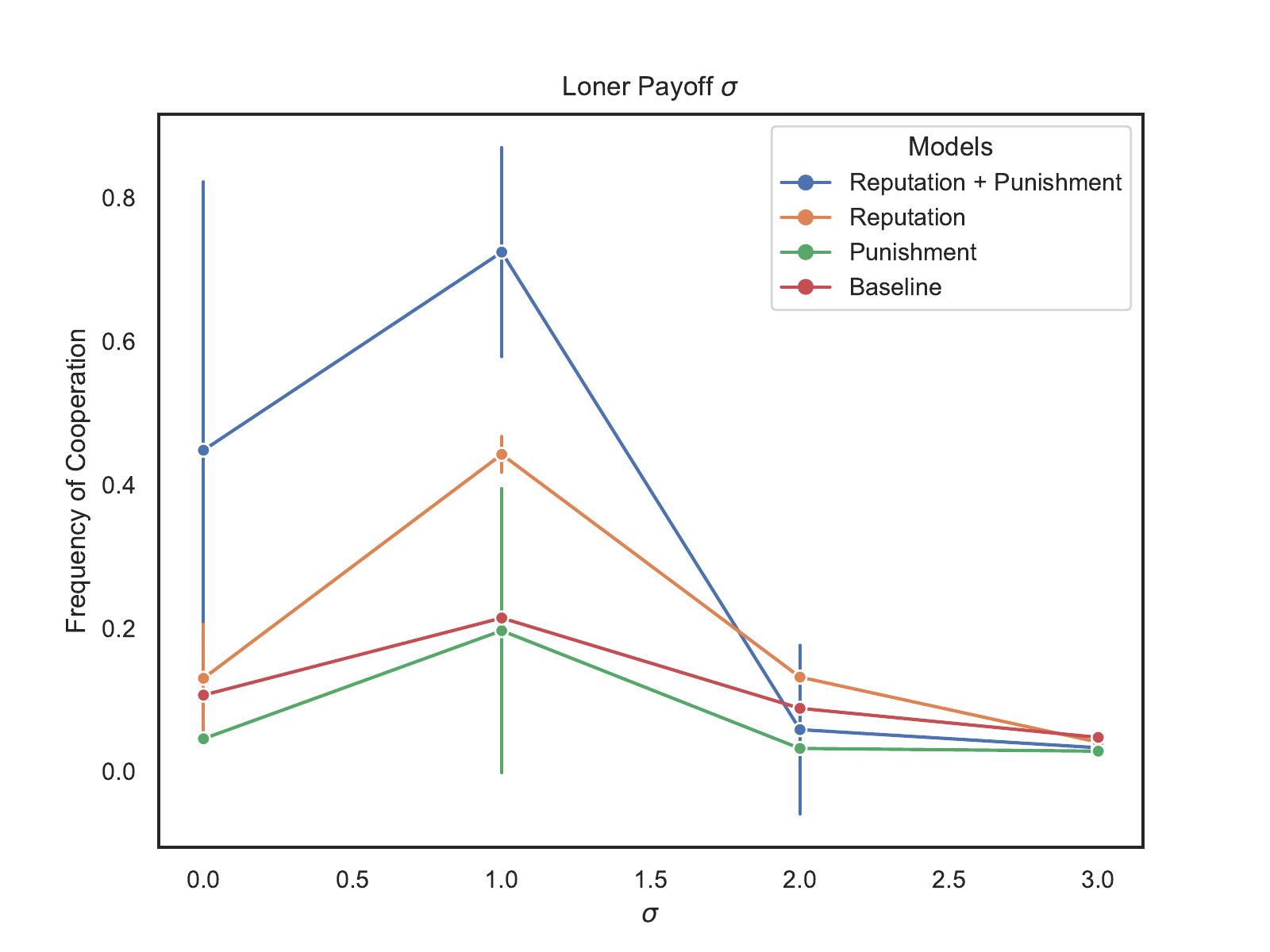}
  \caption{\textbf{Abstaining from the OPGG cannot be too lucrative or cooperation disappears.} We see for all models, the highest levels of cooperation occur when the loner's payoff $\sigma$ is close to 1. Populations that use social norms and punishment are more sensitive towards higher values of $\sigma$, with higher cooperation seen in populations with only reputation. . Results shown use $N=1000, n=5, r=3, \gamma=1, \beta=2, t=2 \times 10^5, \epsilon=0.1, m=0.95, \Omega=\tfrac{10}{11}$, with 25 iterations per data point.}
  \label{fig:sigma}
\end{figure}

\pagebreak
\subsection{Punishment Cost \& Penalty}

\begin{figure}[h]
  \centering
  \includegraphics[width=\linewidth]{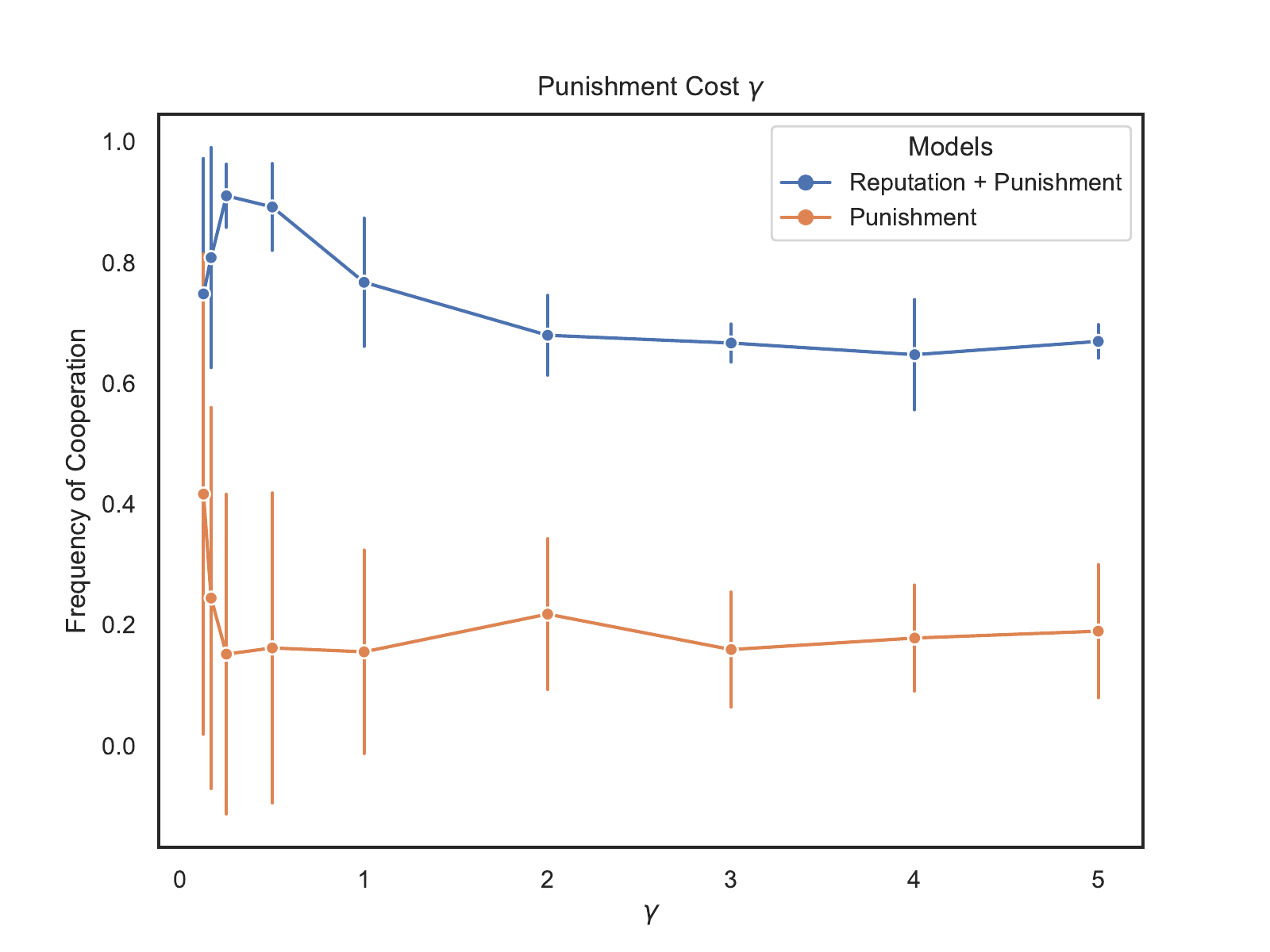}
  \caption{\textbf{Cheaper punishing makes a positive but small difference to cooperation.} When both reputation and punishment is used, as the cost of punishing increases, the level of cooperation observed in the population marginally decreases. When only punishment is used however, for all $\gamma$, cooperation is generally below $40\%$. In the presence of a social norm, the best chance for cooperation lies when the cost of punishment is lower than one, specifically around $\gamma = 0.25$. In the absence of reputation, the cost of punishment needs to be as low as possible to allow for low to moderate levels of cooperation. Results shown use $N=1000, n=5, r=3, \sigma=1, \beta=2, t=2 \times 10^5, \epsilon=0.1, m=0.95, \Omega=\tfrac{10}{11}$.}
  \label{fig:gamma}
\end{figure}
\pagebreak

\begin{figure}[h]
  \centering
  \includegraphics[width=\linewidth]{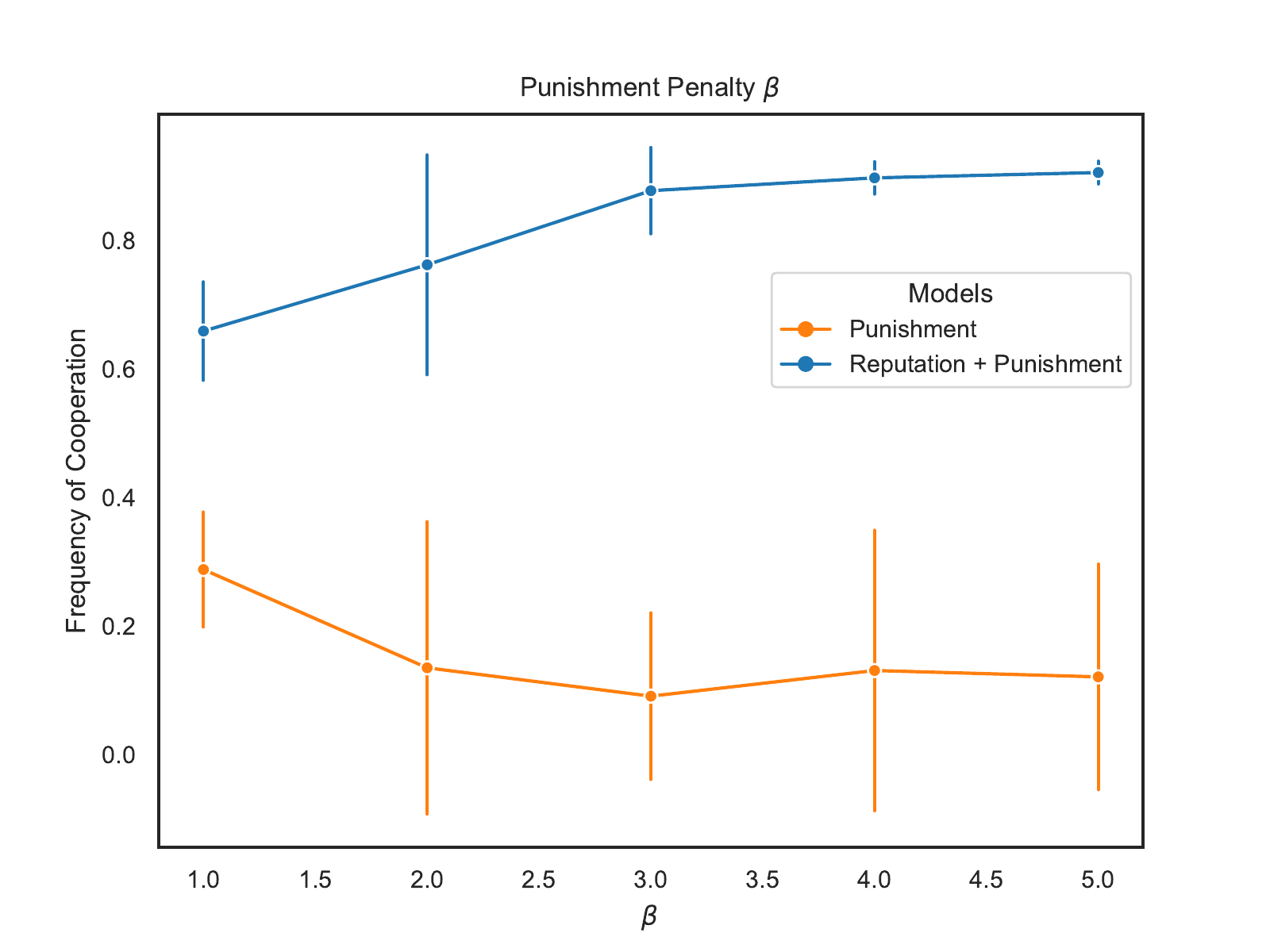}
  \caption{\textbf{Harsher punishments make a positive but small difference to cooperation.} When both reputation and punishment is used, we see a gradual positive trend in the level of cooperation as the punishment penalty increases. When only punishment is used, results are generally stable with low levels of cooperation. Our main findings hold regardless of the harshness of the punishment received. Results shown use $N=1000, n=5, r=3, \sigma=1, \gamma=1, t=2 \times 10^5, \epsilon=0.1, m=0.95, \Omega=\tfrac{10}{11}$.}
  \label{fig:beta}
\end{figure}

\pagebreak
\subsection{Likelihood of further interactions}

\pagebreak
\subsection{OPGG Group Size}

\begin{figure}[!h]
  \centering
  \includegraphics[width=\textwidth]{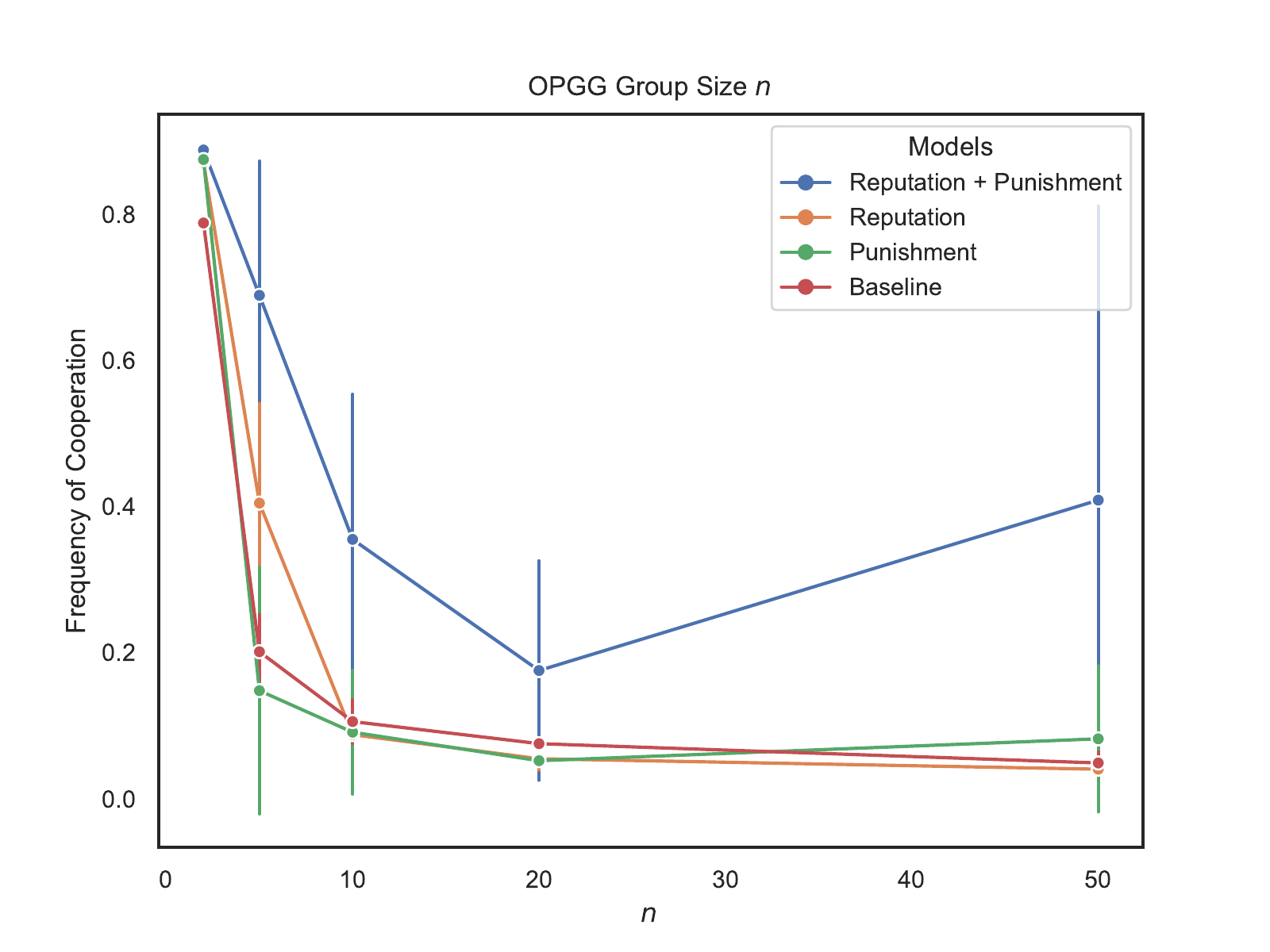}
  \caption{\textbf{Reputation and punishment strengthen cooperation in larger groups while reputation alone does not.} Only when combined with punishment does reputation allow a reasonable chance for cooperation in moderately sized groups. Societies with no punishment but with an Anti-Defector social norm are largely no more or no less effective than the traditional OPGG. Results shown use $N=1000, r=3, \sigma=1, \gamma=1, \beta=2, t=2 \times 10^5, \epsilon=0.1, m=0.95, \Omega=\tfrac{10}{11}$ with 10 iterations per data point.}
  \label{fig:n}
\end{figure}

\pagebreak
\subsection{Degree of Evolutionary Mixing}

\pagebreak
\subsection{Mutation Rate}

\pagebreak

\begin{landscape}
  \section{Table of parameters}
  \begin{table}[h!]
    \centering
    \begin{tabular}{@{}llllllllllllll@{}}
      \toprule
      Figure          & Model                     & Social Norm & $N$  & $n$ & $r$ & $\sigma$ & $t$    & $\gamma$ & $\beta$ & $m$  & $\epsilon$ & $\Omega$         & Repeats \\ \midrule
      1               & *                         & *           & 1000 & 5   & 3   & 1        & 200000 & 1        & 2       & 0.95 & 0.1        & $\tfrac{10}{11}$ & 100     \\
      2               & Reputation                & *           & 1000 & 5   & 3   & 1        & 200000 & 1        & 2       & 0.95 & 0.1        & $\tfrac{10}{11}$ & 100     \\
      3-4             & Reputation and Punishment & *           & 1000 & 5   & 3   & 1        & 200000 & 1        & 2       & 0.95 & 0.1        & $\tfrac{10}{11}$ & 100     \\
      S1-S10, S12-S16 & *                         & *           & 1000 & 5   & 3   & 1        & 200000 & 1        & 2       & 0.95 & 0.1        & $\tfrac{10}{11}$ & 100     \\
      S11             & Baseline \& Punishment    & NA          & 1000 & 5   & 3   & 1        & 200000 & 1        & 2       & 0.95 & 0.1        & $\tfrac{10}{11}$ & 100     \\
      Table S1        & *                         & *           & 1000 & 5   & 3   & 1        & 200000 & 1        & 2       & 0.95 & 0.1        & $\tfrac{10}{11}$ & 100     \\
      S17             & *                         & AD          & 1000 & 5   & *   & 1        & 200000 & 1        & 2       & 0.95 & 0.1        & $\tfrac{10}{11}$ & 25      \\
      S18             & *                         & AD          & 1000 & 5   & 3   & *        & 200000 & 1        & 2       & 0.95 & 0.1        & $\tfrac{10}{11}$ & 25      \\
      S19             & *                         & AD          & 1000 & 5   & 3   & 1        & 200000 & *        & 2       & 0.95 & 0.1        & $\tfrac{10}{11}$ & 10      \\
      S20             & *                         & AD          & 1000 & 5   & 3   & 1        & 200000 & 1        & *       & 0.95 & 0.1        & $\tfrac{10}{11}$ & 25      \\
      S21             & *                         & AD          & 1000 & 5   & 3   & 1        & 200000 & 1        & 2       & 0.95 & 0.1        & *                & 10      \\
      S22             & *                         & AD          & 1000 & *   & 3   & 1        & 200000 & 1        & 2       & 0.95 & 0.1        & $\tfrac{10}{11}$ & 10      \\
      S23             & *                         & AD          & 1000 & 5   & 3   & 1        & 200000 & 1        & 2       & *    & 0.1        & $\tfrac{10}{11}$ & 10      \\
      S24             & *                         & AD          & 1000 & 5   & 3   & 1        & 200000 & 1        & 2       & 0.95 & *          & $\tfrac{10}{11}$ & 25      \\ \bottomrule
    \end{tabular}
    \caption{\textbf{Table of parameters.} An asterisk represents a wildcard indicating multiple values for the respective variable. $N$ is the number of players in the population, $n$ is the size of the group within the OPGG, $r$ is the multiplier on the public good, $\sigma$ is the payoff awarded to those who opt-out of the OPGG, $t$ is the number of periods each simulation lasts, $\gamma$ is the cost to punish another individual, $\beta$ is the financial penalty incurred if punished, $m$ is the degree of evolutionary mixing representing the probability of evolving based on a player outside of the evolving player's group, $\epsilon$ is the probability of a random mutant being introduced into the population, $\Omega$ is the probability of further rounds of the OPGG within the same period. All results in the Supplementary Materials represent averages and standard deviations averaged over a number of repeats. } \label{parameters}
  \end{table}

\end{landscape}

\end{document}